\documentclass[twocolumn,numberedappendix]{aastex62}

\usepackage{times}
\usepackage{graphicx}
\usepackage{epstopdf}
\usepackage{subfigure}
\usepackage{color}
\usepackage{verbatim}
\usepackage{bm}
\usepackage{hyperref}
\usepackage{CJK}

\hypersetup{colorlinks = true, linkcolor = red, anchorcolor = red, citecolor = blue, filecolor = blue, urlcolor = blue, backref = page}

\submitjournal{ApJ}
\shorttitle{Dust Growth and Ring}
\shortauthors{Li et al.}

\begin{document}

\title{Effects of Ringed Structures and Dust Size Growth on Millimeter Observations of Protoplanetary Disks}

\correspondingauthor{Ya-Ping Li}
\email{leeyp2009@gmail.com}

\author[0000-0002-7329-9344]{Ya-Ping Li}
\affil{Theoretical Division, Los Alamos National Laboratory, Los Alamos, NM 87545, USA}
\affil{Shanghai Astronomical Observatory, Chinese Academy of Sciences, 80
Nandan Road, Shanghai 200030, China}
\author[0000-0003-3556-6568]{Hui Li}
\affil{Theoretical Division, Los Alamos National Laboratory, Los Alamos, NM 87545, USA}
\author{Luca Ricci}
\affil{Department of Physics and Astronomy, California State University Northridge, 18111 Nordhoff Street, Northridge, CA 91130, USA}
\author[0000-0002-4142-3080]{Shengtai Li}
\affil{Theoretical Division, Los Alamos National Laboratory, Los Alamos, NM 87545, USA}
\author[0000-0002-1899-8783]{Tilman Birnstiel}
\affil{University Observatory, Faculty of Physics, Ludwig-Maximilians-Universit{\"a}t M{\"u}nchen, Scheinerstr. 1, 81679 Munich, Germany}
\author[0000-0001-8061-2207]{Andrea Isella}
\affil{Department of Physics and Astronomy, Rice University, 6100 Main Street, Houston, TX 77005, USA}
\author[0000-0003-4142-9842]{Megan Ansdell}
\affil{Center for Integrative Planetary Science, Berkeley, CA 94720, USA}
\affil{Department of Astronomy, University of California at Berkeley, Berkeley, CA 94720, USA}
\author[0000-0003-3564-6437]{Feng Yuan}
\affil{Shanghai Astronomical Observatory, Chinese Academy of Sciences, 80
Nandan Road, Shanghai 200030, China}
\affil{School of Astronomy and Space Sciences, University of Chinese Academy of Sciences, No. 19A Yuquan Road, Beijing 100049, China}
\author[0000-0002-9128-0305]{Joanna Dr\c{a}\.{z}kowska}
\affil{University Observatory, Faculty of Physics, Ludwig-Maximilians-Universit{\"a}t M{\"u}nchen, Scheinerstr. 1, 81679 Munich, Germany}
\author[0000-0002-1589-1796]{Sebastian Stammler}
\affil{University Observatory, Faculty of Physics, Ludwig-Maximilians-Universit{\"a}t M{\"u}nchen, Scheinerstr. 1, 81679 Munich, Germany}

\begin{abstract}
{The growth of solids from sub-micron to millimeter and centimeter sizes is the early step 
toward the formation of planets inside protoplanetary disks (PPDs). However, such processes and their potential impact on the 
later stages of solid growth are still poorly understood. 
In this work, we test the hypothesis that most disks contain at least one ringed structure with a relative small radius. 
We have carried out a large family of 1D two-fluid (gas+dust) hydrodynamical simulations 
by evolving the gas and dust motion self-consistently while allowing dust size to evolve via coagulation and fragmentation. 
We investigate the joint effects of ringed structures and dust size growth on the overall sub-millimeter and millimeter (mm) 
flux and spectral index of PPDs.  Ringed structures slow down the dust radial drift and speed up the dust growth. 
In particular, we find that  those unresolved disks with a high fragmentation velocity ($\sim10\ {\rm m~s^{-1}}$) and a high dust surface density ($\sim10\ {\rm g\ cm^{-2}}$ in the ring)  can have mm 
spectral indices as low as $\sim2.0$, consistent with mm observations of faint disks in 
nearby star forming regions. Furthermore, disks with more than one ringed structure can potentially reproduce brighter disks 
with spectral indices lower than $\sim2.5$. 
Future multi-wavelength high-resolution observations of these low spectral index sources can be used to test 
the existence of the ringed structures in the unresolved disks and differentiate the effects of dust size growth from optical depth. 
}
\end{abstract}

\section{Introduction}

The discovery of thousands of exoplanets over the last couple of decades has clearly shown 
that the birth of planets is a very efficient process in nature \citep[e.g.,][]{Burke2015}. 
The commonly accepted theory is that planets form in young disks orbiting pre-main sequence 
(PMS) stars through the agglomeration of small $\mu$m dust particles into km-sized ``planetesimals", 
which are massive enough to gravitationally attract other solids in the disk \citep[see review by][]{Chiang2010}.
Grain growth from sub-micrometer sizes, which are the typical sizes of dust in the 
interstellar medium (ISM), to millimeter and centimeter sizes is thus the first step toward 
the formation of planetesimals inside young circumstellar disks.
However, the formation of planetesimals is still poorly understood due to \textit{i}) the low 
sticking efficiency of grains larger than $\sim 1$ mm-cm, and \textit{ii}) the mutual dynamical 
interaction between dust grains and the gas in the disk.  In the case of a gas-rich disk with 
density and temperature both decreasing with distance
from the central star, small solids radially drift inward because of the aerodynamic drag by
the gas orbiting at sub-Keplerian speeds \citep[][]{Weidenschilling1977,Brauer2007}. 
Models of the evolution of disk solids have
predicted radial drift timescales which are too short ($100\sim1000$ orbits) to 
form planetesimals \citep[see review by][]{Johansen2014}. Understanding dust coagulation  might be of fundamental importance to understand the formation of planetesimals.

Once dust particles grow from ISM-like sizes of $\mu$m to mm/cm sizes, they can be 
mainly traced by their continuum emission at sub-millimeter (sub-mm) and longer wavelengths. 
Therefore, sub-mm observations have the potential of setting strong constraints on the initial 
stages of the planet formation process.
In the last two decades, several observations of protoplanetary disks (PPDs) at sub-mm/mm 
wavelengths have reported relatively shallow slopes with values of the mm spectral index 
$2.0 \leq \alpha_{\rm 1-3mm} \leq 3.0$ ($F_{\nu}\propto\nu^{\alpha}$). These values are 
significantly lower than those measured for dust in the ISM, and are taken as observational 
evidence for dust growth to mm-sized grains in disks under the assumption that the dust 
emission is mostly optically thin 
(e.g., \citealt{Testi2003,Natta2004,Rodmann2006,Ricci2010a,Pinilla2014,Andrews2015,Ansdell2018}). 
\citet{Ricci2012} explored the effect of local optically thick emission 
regions in the mm emission without mm/cm grains.
They found that optically thick regions with relatively small filling factors 
can reproduce the mm spectral indices of disks without grain growth, although the 
dust over-densities required by the optically thick effect are generally larger than those predicated by the typical known physical processes proposed in the literature.
These processes include vortex formation, streaming instabilities, disk 
viscosity transitions \citep[e.g.,][]{Lovelace1999,Li2000,Li2001,Klahr2003,Rice2004,Fromang2005,Johansen2007,Johansen2009,Boss2010,Regaly2012}.

\citet{Birnstiel2010b} have tested the grain growth model 
by millimeter observations of dust continuum, and they found that it can naturally reproduce the 
observed mm-slope if the radial drift of the dust is halted by an unknown mechanism. 
It is generally believed that some inhomogeneities in the gas density, either azimuthally symmetric, 
such as rings, or with azimuthal asymmetries, such as vortices, have to be invoked to slow down 
the dust radial drift because of the gas pressure structure \citep[e.g.,][in the case of 
dust trapping by azimuthally symmetric rings]{Pinilla2012}.    The dust can, therefore, 
be trapped effectively. Grains can collide frequently and stick together by van der Waals forces, 
which leads to the formation of large-size dust particles 
\citep[e.g.,][]{Dominik1997,Poppe2000,Blum2008,Brauer2008,Birnstiel2010a}.
Such dust accumulation regions can appear as bright rings in PPDs observed from 
the dust continuum emission at millimeter (mm) wavelengths.  
There is now strong observational evidence for the existence of multiple rings (or rings sandwiched by gaps) 
in disks by high-resolution observations with the Atacama Large Millimeter Array (ALMA), 
e.g., HL Tau \citep{ALMA2015}, TW Hya \citep{Andrews2016,Tsukagoshi2016}, HD 163296 \citep{Isella2016}, 
AA Tau \citep{Loomis2017}, Elias 24 \citep{Cieza2017,Cox2017,Dipierro2018}, AS 209 \citep{Fedele2018}, 
GY 91 \citep{Sheehan2018}, V1094 Sco \citep{Ansdell2018,Terwisga2018}, MWC 758 \citep{Boehler2018}, 
more disks in a recent survey of young disks in Taurus \citep{Long2018}, 20 disks from the Disk Substructures at High Angular Resolution Project (DSHARP) observed very recently \citep{Andrews2018DSHARP,Dullemond2018,Huang2018DSHARP}, and 16 disks from ALMA archival data \citep{vanderMarel2019}. These high-resolution observations have also revealed that these narrow rings have relatively high optical depth at mm band \citep[e.g.,][]{ALMA2015,Tsukagoshi2016,Andrews2018DSHARP,Dullemond2018,Isella2016,Isella2018,Huang2018,Huang2018DSHARP}. 
\citet{Dullemond2018} find that the narrow radial size of the rings strongly supports the 
dust trapping scenario.  Such dust traps can also facilitate the grain growth.

\begin{figure}[htbp]
\begin{center}
\includegraphics[width=0.45\textwidth]{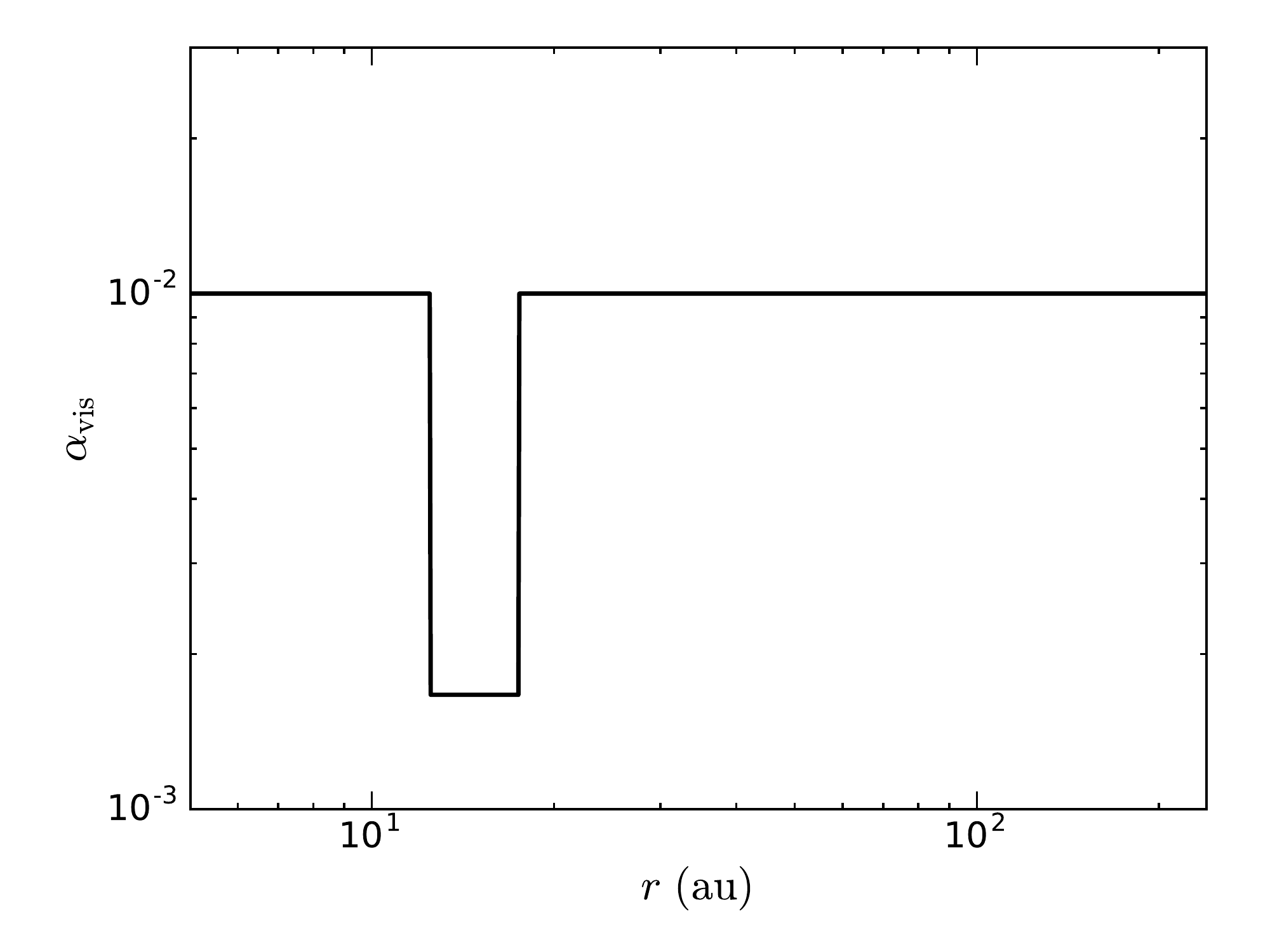}
\end{center}
\caption{The assumed disk viscosity radial profile $\alpha_{\rm vis}$  for most of our models.}\label{fig:viscosity}
\end{figure}

Motivated by the fact that dust rings appears to be ubiquitous at least among the disk 
sample observed so far, we explore their consequences on the overall (sub-)mm disk emission properties. 
For the sake of simplicity, we use a non-monotonic disk viscosity profile (e.g., Figure~\ref{fig:viscosity})
to produce a local gas surface density bump. We perform detailed two-fluids (gas+dust) hydrodynamics
simulations, in which the dust coagulation and disk hydrodynamics evolution are treated together, 
and calculate the dust continuum emission at several mm bands with detailed radiative transfer.
Our main goal is to interpret the low spectral indices observed for many young disks, 
and understand the link between the spectral slope and the dust dynamics in the disks.
Based on the comparison between our simulations and the observed dust emission, 
we can constrain some fundamental parameters for the dust coagulation model and provide 
new insight into the mechanism of dust trapping and growth.

The paper is organized as follows. The numerical simulation method adopted in this work is 
presented in  Section~\ref{sec:method}. We present the results in Section~\ref{sec:results}, 
and summarize the main results and 
discuss the physical implications in Section~\ref{sec:conclusions}.

\section{Model} \label{sec:method}

We envision the following scenario: the disk gas surface density profile contains at least one ring in the inner disk region 
(which is unresolved for most disks currently).
By performing 1D hydrodynamical simulations using \texttt{LA-COMPASS} \citep{Li2005,Li2009,Fu2014}, 
we study the gas + dust dynamics spanning several million years, 
including dust coagulation and fragmentation self-consistently 
\citep{Birnstiel2010a}, and then produce the dust emission at different mm wavelengths 
by making use of the \texttt{RADMC-3D} package \citep{Dullemond2012} to compare with observations.

\subsection{Hydrodynamics and Model Assumption}

Without loss of generality, we consider a disk around a PMS star with a mass of $0.6~M_{\sun}$ 
at a distance of 140 pc. The disk extends from 5 au to 240 au. Disk self-gravity in gas and dust is not included. 
To create dust rings in such disks, we use varying disk viscosity radial profiles to achieve such configurations. 
The gas viscosity is adopted from the $\alpha$-prescription $\nu_{\rm g}=\alpha_{\rm vis} c_{\rm s}h_{\rm g}$ 
with $h_{\rm g}$ being the gas disk height and $\alpha_{\rm vis}$ the dimensionless 
viscosity parameter \citep{Shakura1973}. For simplicity, 
in our fiducial model, $\alpha_{\rm vis}(r)=\alpha_{0}=10^{-2}$ for the most region of the disk. 
We also assume a rectangular gap with a low viscosity $\alpha_{\rm vis}$ in its radial profile.  
The viscosity gap is centered at 15 au with a gap width $w_{\rm ring}=5\ {\rm au}$ and 
a value for the viscosity in the dip $\alpha_{\rm min}=\alpha_{0}/(1+d_{\rm ring})=1.7\times10^{-3}$, 
which is shown in Figure~\ref{fig:viscosity}.
As we will see, the viscosity gap can produce a gas bump, which can then slow down the 
radial drift of dust particles in the disk, creating a dust ring.

The existence of a dust ring in the disk, though a hypothesis in this paper, is well motivated by current high-resolution observations mentioned above. Actually, any mechanism that can produce a similar gas bump, even without invoking a viscosity jump, could produce results similar to the ones discussed here.
Both viscosity transition \citep[e.g.,][]{Varniere2006} and the presence of a planet \citep[e.g.,][]{Bryden1999,Papaloizou2007,Dipierro2015,Jin2016,Dong2017,Liu2018,Zhang2018} could lead such a dust ring configuration. 
The viscosity transition can be characterized by a gap in the viscosity around a few au (e.g., see review by \citealt{Armitage2011}), which is slightly smaller than the radius of our viscosity gap. However, the inner edge and the radial width of the dead zone are actually not well constrained and will depend in general on the ionization sources and the physical structure of the disk, as well as on the intensity of the magnetic field in the disk itself. Given these uncertainties, it makes sense to explore a broad range of values for the central radius and width of the viscosity jump, as will be done in this work. 
For the planet-disk interaction scenario, some kinematic evidence of protoplanets forming within ringed-structure disks have been found recently based on the detection of a localized deviation from Keplerian velocity in HD 163296 \citep{Pinte2018,Teague2018}, although other possibilities (e.g., zonal flow) for the inner ringed-structures cannot be fully ruled out yet.

The initial gas surface density profile $\Sigma_{\rm g}(r)$ is set as follows:
\begin{equation}\label{eq:gas}
  \Sigma_{\rm g}(r)=\Sigma_{0}\left(\frac{r}{r_{\rm c}}\right)^{-\gamma}\exp\left[-\left(\frac{r}{r_{\rm c}}\right)^{2-\gamma}\right],
\end{equation}
where $r_{\rm c}=60~\rm au$ and $\gamma=1$ unless otherwise stated. 
These parameters represent a typical disk as seen from observations \citep{Andrews2010}. 
The normalization of gas surface density $\Sigma_{0}$ is treated as a parameter listed in 
Table~\ref{tab:para}. The locally isothermal sound speed $c_{\rm s}$ is chosen as
\begin{equation}\label{eq:cs}
  \frac{c_{\rm s}}{v_{\rm K}}=h_{0}\left(\frac{r}{r_{0}}\right)^{0.25},
\end{equation}
where $v_{\rm K}(r)$ is the local Keplerian velocity, $r_{0}=5$ au unless otherwise stated. 
$h_{0}$ is listed in Table~\ref{tab:para} with a typical value of $0.04$. 
This corresponds to a disk temperature profile as $T\propto r^{-0.5}$. 
Equation~(\ref{eq:cs}) also expresses the radial profile of the gas scale height $h_{\rm g}/r$.  
The gas and dust fluids are evolved following the conservation of mass, radial, 
and angular momentum equations.

The continuity equations of gas is
\begin{equation}\label{eq:mass_gas_cons}
\frac{\partial \Sigma_{\rm g}}{\partial t} + \nabla \cdot (\Sigma_{\rm g} \mathbf{v}_{\rm g}) =0,
\end{equation}
where $\mathbf{v}_{\rm g}$ is the gas fluid velocity.
The momentum equation for the gas is
\begin{equation}
\frac{\partial (\Sigma_{\rm g}\mathbf{v}_{\rm g})}{\partial t} +\nabla(\mathbf{v}_{\rm g}\cdot\Sigma_{\rm g} \mathbf{v}_{\rm g})+\nabla P=-\Sigma_{\rm g}\nabla\Phi_{\star}+\Sigma_{\rm g} \mathbf{f}_{\nu}-\Sigma_{\rm d} \mathbf{f}_{\rm d},
\label{eq:momen_gas_cons}
\end{equation}
where $\bf{f_{\nu}}$ indicates the viscous force from the Shakura-Sunyaev disk \citep{Shakura1973}. 
We adopt an equation of state $P=c_{\rm s}^{2}\Sigma_{\rm g}$ for the gas component, 
where $P$ is the vertically integrated gas pressure.  $\Phi_{\star}$ is the gravitational potential 
of the central star, $\mathbf{f}_{\rm d}$ is the summation of $\mathbf{f}_{\rm d}^{i}$ over $i$ 
and $\mathbf{f}_{\rm d}^{i}$ is the drag force between the gas and dust species $i$, which is defined as
\begin{equation}\label{eq:drag}
\mathbf{f}_{\rm d}^{i}=\frac{\Omega_{\rm k}}{{\rm St}^{i}}(\mathbf{v}_{\rm g}-\mathbf{v}_{\rm d}^{i}),
\end{equation}
where $\Omega_{\rm k}$ is the Keplerian angular velocity. ${\rm St^{i}}$ is the 
Stokes number of the dust species $i$. We can estimate this feedback term based on \citet{Takeuchi2002}. 
In the Epstein regime for the disk and dust parameters of interest here, the Stokes number of the particles with dust radius $a$ in the mid-plane of the disk is defined as 
\begin{equation}\label{eq:st}
  {\rm St}^i = \frac{\pi\rho_{\rm s}a^i}{2\Sigma_{\rm g}}.
\end{equation}
The dust size corresponding to ${\rm  St}=1$ is thus
\begin{equation}\label{eq:amax_st}
  a_{\rm St=1}=\frac{2\Sigma_{\rm g}}{\pi\rho_{\rm s}}.
\end{equation}
The Stokes number in the Epstein regime can then be re-written as ${\rm St}^i=a^i/a_{\rm St=1}$.

We treat the dust component as a pressure-less fluid. The dust feedback, i.e., drag forces between the gas and dust, are incorporated into the 
momentum equation for both the gas and dust \citep{Fu2014}. 
The continuity equation of dust species $i$ is:
\begin{equation}\label{eq:mass_dust_cons}
\frac{\partial \Sigma_{\rm d}^{i}}{\partial t} +\nabla \cdot (\Sigma_{\rm d}^{i} \mathbf{v}_{\rm d}^{i}) = \nabla \cdot \left(\Sigma_{\rm g} D_{\rm d}^{i} \nabla \left(\frac{\Sigma_{\rm d}^{i}}{\Sigma_{\rm g}}\right)\right).
\end{equation}
Here $D_{\rm d}$, which describes the dust diffusivity, is related with the Stokes number 
${\rm St}$  as $D_{\rm d}=\nu_{\rm g}/(1+{\rm St}^{2})$ \citep{Youdin2007}, 
$\nu_{\rm g}$ is the gas viscosity. $\Sigma_{\rm d}^{i}$, $\mathbf{v}_{\rm d}^{i}$ are 
dust surface density and velocity for species $i$, respectively. 
The momentum equation for 
dust species $i$ after including the drag force from the gas is
\begin{equation}
\frac{\partial \Sigma_{\rm d}^{i}\mathbf{v}_{\rm d}^{i}}{\partial t} +\nabla(\mathbf{v}_{\rm d}^{i}\cdot\Sigma_{\rm d}^{i} \mathbf{v}_{\rm d}^{i})=-\Sigma_{\rm d}^{i}\nabla\Phi_{\star}+\Sigma_{\rm d}^{i} \mathbf{f}_{\rm d}^{i}. 
\label{eq:momen_dust_cons}
\end{equation}
For an illustrative purpose, the radial velocity of the dust can be estimated as 
\begin{equation}\label{eq:vr_dust}
v_{r,{\rm d}}=\frac{{\rm St}^{-1}v_{r,{\rm g}}-\eta v_{\rm K}}{{\rm St}^{-1}+{\rm St}},
\end{equation}
where $\eta=-\frac{c_{\rm s}^2}{v_{\rm K}^2}\frac{d\ln P}{d\ln r}$, and
\begin{equation}\label{eq:vr_gas}
v_{r,{\rm g}}\simeq-\frac{3}{\Sigma_{\rm g}\sqrt{r}}\frac{\partial}{\partial r}(\Sigma_{\rm g }\nu_{\rm g}\sqrt{r}),
\end{equation}
is the radial velocity of the gas. Here we omit the superscript of $i$ for different dust species. 
We can see that the dust radial velocity is composed of two terms, 
namely gas drag (the first term) and radial drift (the second one) in Equation~(\ref{eq:vr_dust}). 
Then the drag force between dust and gas for each dust species 
can be obtained by combining Equations~(\ref{eq:drag}$,$\ref{eq:vr_dust}$-$\ref{eq:vr_gas}).

To model the dust size growth during disk evolution, we calculate the evolution of dust size distribution 
which is discretized logaritmically between $1\ \mu{\rm}$m and $100\ {\rm cm}$. We use 25 bins per size decade, 
which leads to 151 dust species with their size $a_{i}$. Such a dust size resolution can avoid an artificial dust size growth due to some numerical diffusion in the Smoluchowski algorithm \citep{Smoluchowski1916}, and ensure the convergence 
of dust growth rate \citep{Ohtsuki1990}.

The dust size evolution is computed within each spatial cell and is 
handled via an operator splitting approach
between hydrodynamics and dust coagulation/fragmentation. 
Due to the high computational cost of solving dust coagulation, we implement a sub-stepping routine, 
where the dust coagulation module is executed every 50 time-steps of the hydrodynamical simulations, 
which still gives good calculation accuracy for the coagulation algorithm.
The dust coagulation and fragmentation model follows \citet{Birnstiel2010a} with an explicit integration 
scheme by solving the Smoluchowski equation, which includes the radial drift and turbulent mixing as the 
sources of collision velocities. 
Turbulence is the major source of collision velocity. Since this velocity increases with 
Stokes number, we can derive the maximum size of grains that can grow before the 
impact velocity exceeds the fragmentation threshold velocity $v_{\rm f}$ as \citep{Birnstiel2012,Pinilla2012}
\begin{equation}\label{eq:amax_frag}
  a_{\rm max}=\frac{4\Sigma_{\rm g}}{3\pi\alpha_{\rm vis}\rho_{\rm s}}\frac{v_{\rm f}^{2}}{c_{\rm s}^{2}},
\end{equation}
where $\rho_{\rm s}$ is the solid density of the dust particles. Note that this 
$a_{\rm max}$ is only valid for grains with ${\rm St}\leqslant1$. This size limit for dust coagulation 
is also referred as the fragmentation barrier. The inverse dependence on viscosity parameter is because the turbulent velocity depends on $\alpha_{\rm vis}$.

We adopt the following procedure in simulations: 
Initially, only 1 $\mu$m sized dust particles are included in the disk, and the surface density distribution 
follows the radial profile of the gas with an initial radial-independent dust-to-gas mass ratio of $0.01$. 
The evolution for each dust species $a_{i}$ and the gas component is determined by the mass and 
momentum equations described above. We can self-consistently consider the 
hydrodynamics of both the gas and multiple dust species with the full dust coagulation and 
fragmentation. Previous studies by \citet{Pinilla2016} (see also \citealt{Birnstiel2010a,Pinilla2012,Pinilla2014}) have produced very interesting results on the effects of dust size growth
during the disk evolution. In our current study, we treat the gas-dust dynamical interaction self-consistently, which goes beyond some of the previous work where the gas evolution is not considered \citep{Pinilla2012,Pinilla2014}. 
\citet{Birnstiel2010a} (see also \citealt{Pinilla2016}) has considered the viscous evolution 
of the gas, but without taking into account the full coupling (dust feedback) between the gas and dust.

Unless otherwise stated, we solve the 1D hydrodynamics equations with a logarithmically 
radial grid of $n_{r}=1024$. 
An outflow inner boundary condition and outer boundary condition are imposed on the dust and gas, 
which allow the gas/dust flow out and flow in from the boundary depending on their radial velocity.

\subsection{Radiative Transfer}\label{sec:radmc}

After having obtained the dust surface density distribution in disks, the dust continuum emission 
at different mm wavelengths was computed using the \texttt{RADMC-3D} package \citep{Dullemond2012}.
A two-stage procedure is adopted for modeling the dust continuum radiative process. 
We first convert the 1D dust surface density produced from hydrodynamical simulation into a 3D 
distribution by assuming a dust scale height $h_{\rm d}(r)=0.1h_{\rm g}(r)$ with azimuthal 
symmetry as adopted in previous works \citep[e.g.,][]{Isella2016,Liu2018}\footnote{For our 
fiducial model presented below, we have also tested a different dust scale height $h_{\rm d}=h_{\rm g}\min\left({1,\sqrt{\frac{\alpha_{\rm vis}}{\min(0.5,\rm St)(1+\rm St^2)}}}\right)$ 
by considering dust vertical settling \citep{Birnstiel2010a}, which does not show a significant 
impact on the dust emission in terms of our interest. Specifically, it will lead to an increase of 
dust mm flux of $\sim20\%$, and a slight change of the spectral index ($\sim0.2$).}. 
The vertical density structure is simply scaled as 
$\frac{\Sigma_{\rm d}}{\sqrt{2\pi} h_{\rm d}}\exp(-z^2/2h_{\rm d}^{2})$. 
We use 300 grids along the radial direction with a grid refinement around the bump and 
40 uniform grids in the $\theta$ direction between $70^{\circ}-90^{\circ}$ with 
a mirror symmetry in the equatorial plane. This grid can recover all the dust mass from 
1D hydrodynamics within a $5\%$ uncertainty. A larger grid number does not change the results.

\begin{figure}[htbp]
\begin{center}
\includegraphics[width=0.45\textwidth]{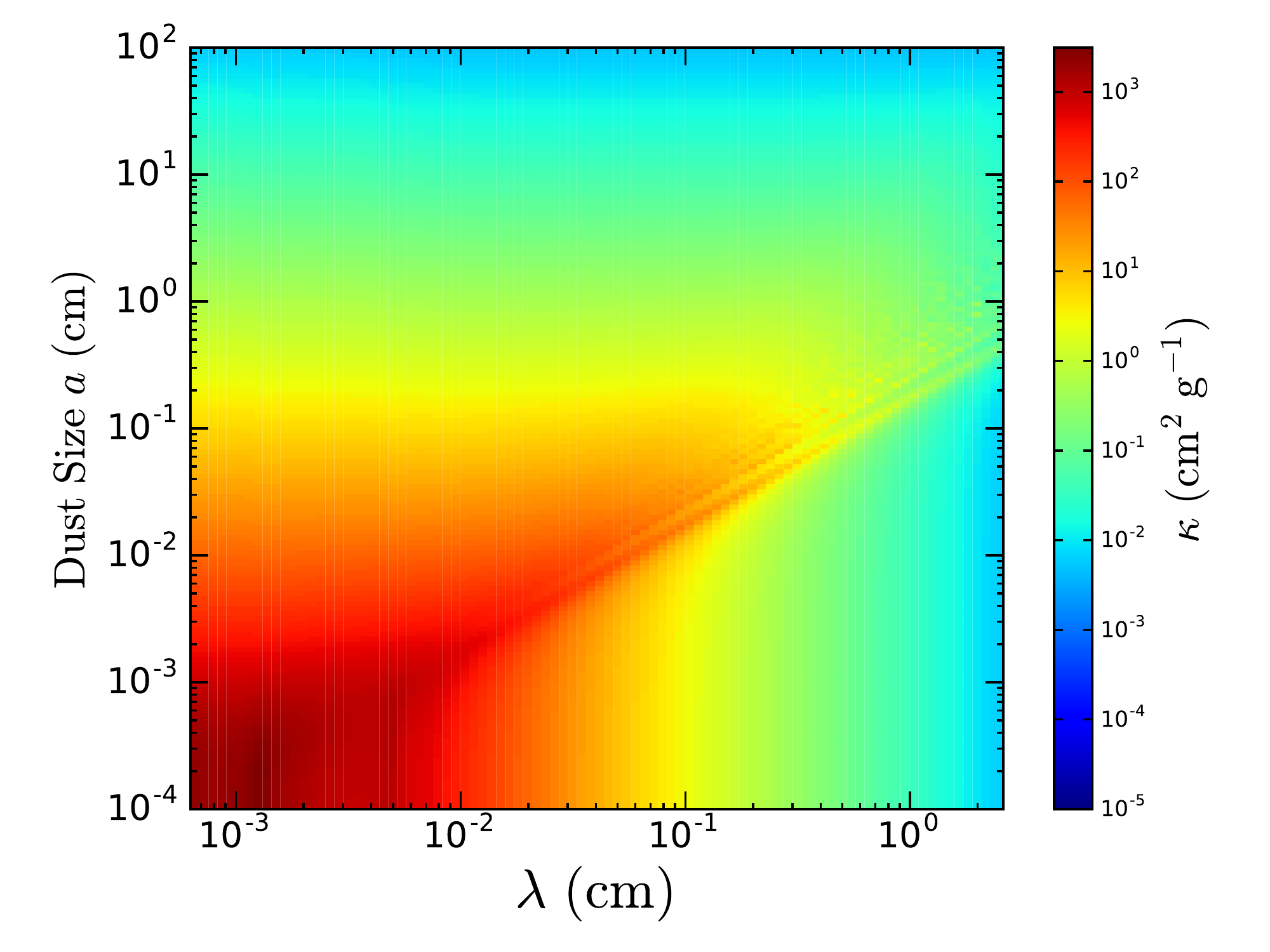}
\hskip 0.0truecm
\includegraphics[width=0.45\textwidth]{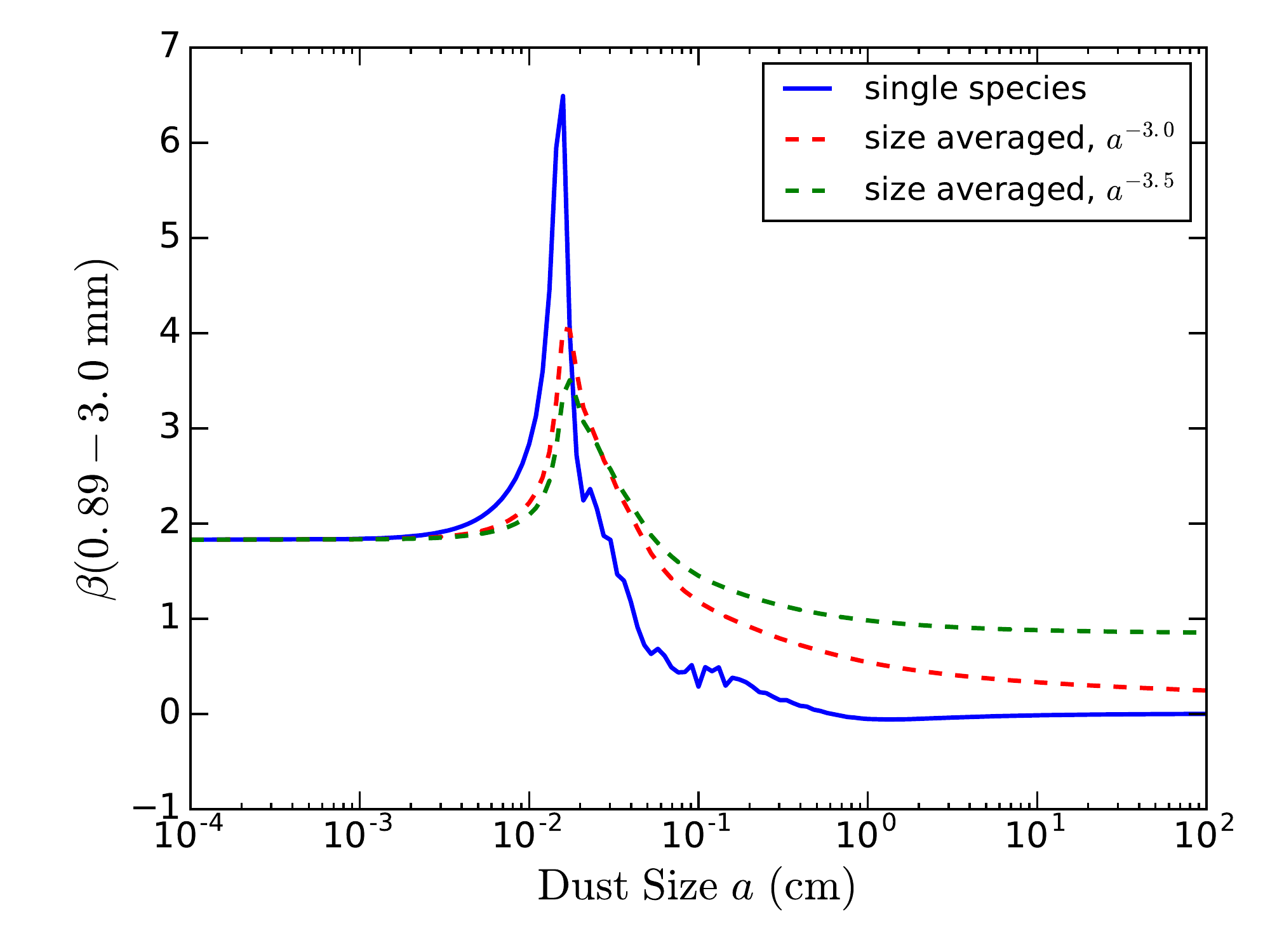}
\end{center}
\caption{Dust opacity for different grain sizes. The upper panel represents the dust opacity $\kappa$ as a function of wavelength and dust size, the lower panel shows the opacity spectral index $\beta$ between 0.89 and 3.0 mm as a function of a single species grain size (solid line).  For a comparison, we also show the $\beta$ profile for the size averaged opacity with a power-law distributed grain (dashed lines).}\label{fig:opacity}
\end{figure}

We then ran \texttt{RADMC-3D} simulations to compute the dust temperature $T_{\rm d}(r,z)$ 
contributed by the stellar radiation. For simplicity, the dust temperature for all 
dust species is assumed to be the same, and controlled by the micro-sized dust particles.\footnote{Dust particles with different sizes should have different temperatures, with the smaller ones being slightly hotter due to a larger scale height. However, such effect should be minor since most of the emission comes from the region at the mid-plane.} The dust opacity coefficient, 
which depends on grain sizes, chemical compositions, and shapes, is another 
important quantity for the modelling. We adopted the dust opacity as a function of 
wavelength and dust size from \citeauthor{Isella2009} (\citeyear{Isella2009}) and \citealt{Ricci2010a}. 
Figure~\ref{fig:opacity} presents the details of this model.  We will also discuss how a 
different dust opacity can effect our results in the following sections \citep{Semenov2003}. 
The dust opacity is dominated by the diagonal region with $\lambda\simeq2\pi a$ 
and $\kappa$ decreases with $a$.

With the dust temperature $T_{\rm d}$ obtained from \texttt{RADMC-3D} and the dust surface 
density $\Sigma_{\rm d}$ for each dust species derived from hydrodynamical simulations, we can calculate the continuum emission for each dust species by ray-tracing with \texttt{RADMC-3D}.
The surface brightness of the continuum emission can be approximated as
\begin{equation}\label{eq:intensity}
  I(r)=B_{\nu}\left(T_{\rm d}(r)\right)\left(1-e^{-\tau_{\nu}(r)}\right),
\end{equation}
where the optical depth $\tau_{\nu}(r)=\kappa_{\nu}\Sigma_{\rm d}(r)/\cos \theta$, 
$B_{\nu}$ is the Planck function, $\theta$ is the disk inclination. 
We assume the disk inclination angle of $\theta=45^\circ$ and a position angle 
of $-30^\circ$ for all our models without losing generality.
Using these quantities, we compute the corresponding spectral energy distribution 
and the radial profile of the surface brightness at several wavelengths between 
$0.89$ mm and $7.0$ cm by \texttt{RADMC-3D}.

In regions of the disk where the emission at a given wavelength is optically thick, 
the dust continuum flux and slope are fully determined by $B_{\nu}(T_{\rm d}(r))$, 
which results in a spectral slope of $2.0$ in the Rayleigh-Jeans limit. 
Conversely, in the optically thin regime the dust emission is related to the opacity coefficient as 
$F_{\nu}\propto\kappa_{\nu}B_{\nu}$. If the emission is in the Rayleigh-Jeans limit, 
then $F_{\nu}\propto\nu^{2+\beta}$, where $\beta$ is the slope of the dust opacity 
coefficient ($\kappa_{\nu}\propto\nu^{\beta}$). $\beta$ changes as dust grows bigger. 
$\beta$ can approach to zero once dust particle grows to size larger than a few mm, 
as seen in the lower panel of Figure~\ref{fig:opacity}. This suggests that dust growth 
can be an important factor for setting the spectral shape of the dust continuum in the 
optically thin regions of the disk \citep[e.g.,][]{Wilner2000,Testi2001,Testi2003,Wilner2005,Draine2006}.

\begin{table*}[htbp]
  \begin{center}
  \caption{\bf Model parameters and simulation results}\label{tab:para}
  \begin{tabular}{lcccccccccccc}
     \hline\hline
     model & $\Sigma_{0}$ & $v_{\rm f}$ & $\alpha_{0}$ & $h_{0}$ & $\gamma$  & $r_{\rm ring}$ & $w_{\rm ring}$ & $d_{\rm ring}$ & $F_{\rm 1mm}$ & $\alpha_{\rm 0.8-1.3{\rm mm}}$ & $\alpha_{\rm 1.0-3.0{\rm mm}}$ & $\alpha_{\rm 3.0-7.0{\rm mm}}$\\
                & $(\rm g~cm^{-2})$ & $({\rm cm~s^{-1}})$ &  &  &   & $(\rm au)$ & $(\rm au)$ &  & $(\rm mJy)$ &  &  & \\
     \hline
     m0v1 & 12.8 & $3\times10^{3}$ & $1\times10^{-2}$& 0.04  & 1.0 & 15.0 & 5.0 & 5.0 & 4.79 & 2.92 &  3.01 & 2.41\\
     m1v1 & 4.3 & $3\times10^{3}$ & $1\times10^{-2}$& 0.04  & 1.0 & 15.0 & 5.0 & 5.0 & 1.52 & 3.00 &  3.00 & 2.30\\
     m2v1 & 0.43 & $3\times10^{3}$ & $1\times10^{-2}$& 0.04  & 1.0 & 15.0 & 5.0 & 5.0 & 0.11 & 2.67 &  2.62 & 2.12\\
     m3v1 & 0.043 & $3\times10^{3}$ & $1\times10^{-2}$& 0.04  & 1.0 & 15.0 & 5.0 & 5.0 & 0.005 & 2.41 &  2.36 & 2.08\\
     m0v2 & 12.8 & $1\times10^{3}$ & $1\times10^{-2}$& 0.04  & 1.0 & 15.0 & 5.0 & 5.0 & 51.5 & 2.38 &  2.44 & 2.42\\
     m1v2 (fiducial) & 4.3 & $1\times10^{3}$ & $1\times10^{-2}$& 0.04  & 1.0 & 15.0 & 5.0 & 5.0 & 32.5 & 2.43 &  2.47 & 2.33\\
     m2v2 & 0.43 & $1\times10^{3}$ & $1\times10^{-2}$& 0.04  & 1.0 & 15.0 & 5.0 & 5.0 & 15.7 & 2.24 &  2.35 & 4.10\\
     m3v2 & 0.043 & $1\times10^{3}$ & $1\times10^{-2}$& 0.04  & 1.0 & 15.0 & 5.0 & 5.0 & 3.12 & 3.86 &  3.77 & 3.53\\
     m0v3 & 12.8 & $3\times10^{2}$ & $1\times10^{-2}$& 0.04  & 1.0 & 15.0 & 5.0 & 5.0 & 69.7 & 2.69 &  2.67 & 2.71\\
     m1v3 & 4.3 & $3\times10^{2}$ & $1\times10^{-2}$& 0.04  & 1.0 & 15.0 & 5.0 & 5.0 & 44.2 & 2.52 &  2.55 & 3.98\\
     m2v3 & 0.43 & $3\times10^{2}$ & $1\times10^{-2}$& 0.04  & 1.0 & 15.0 & 5.0 & 5.0 & 9.32 & 3.52 &  3.53 & 3.51\\
     m3v3 & 0.043 & $3\times10^{2}$ & $1\times10^{-2}$& 0.04  & 1.0 & 15.0 & 5.0 & 5.0 & 1.43 & 3.55 &  3.57 & 3.50\\
      m0v4 & 12.8 & $1\times10^{2}$ & $1\times10^{-2}$& 0.04  & 1.0 & 15.0 & 5.0 & 5.0 & 69.9 & 2.80 &  2.91 & 3.50\\
     m1v4 & 4.3 & $1\times10^{2}$ & $1\times10^{-2}$& 0.04  & 1.0 & 15.0 & 5.0 & 5.0 & 34.4 & 3.27 &  3.33 & 3.49\\
     m2v4 & 0.43 & $1\times10^{2}$ & $1\times10^{-2}$& 0.04  & 1.0 & 15.0 & 5.0 & 5.0 & 7.41 & 3.41 &  3.45 & 3.50\\
     m3v4 & 0.043 & $1\times10^{2}$ & $1\times10^{-2}$& 0.04  & 1.0 & 15.0 & 5.0 & 5.0 & 0.90 & 3.56 &  3.58 & 3.48\\
    ...  &...  &...   &... &...              &...  &...  &...  &...   & ... & ... & ...& ...\\
     m1a1 & 4.3 & $1\times10^{3}$ & $5\times10^{-3}$& 0.04  & 1.0 & 15.0 & 5.0 & 5.0 & 23.1 & 2.30 &  2.34  & 2.36\\
     m1a2 & 4.3 & $1\times10^{3}$ & $2\times10^{-3}$& 0.04  & 1.0 & 15.0 & 5.0 & 5.0 & 4.75 & 3.05 &  3.16 & 2.85\\
     m1a3 & 4.3 & $1\times10^{3}$ & $5\times10^{-4}$& 0.04  & 1.0 & 15.0 & 5.0 & 5.0 & 6.76 & 2.99 &  3.11 & 3.10\\
     ...  &...  &...   &... &...              &...  &...  &...  &...   & ... & ... & ...& ...\\
     m1d1 & 4.3 & $1\times10^{3}$ & $1\times10^{-2}$& 0.04  & 1.0 & 15.0 & 5.0 & 2.0 & 29.8 & 2.82 &  2.97 & 3.40\\
     m1d2 & 4.3 & $1\times10^{3}$ & $1\times10^{-2}$& 0.04  & 1.0 & 15.0 & 5.0 & 10.0 & 23.7 & 2.28 &  2.31 &2.31\\
     m1d3 & 4.3 & $1\times10^{3}$ & $1\times10^{-2}$& 0.04  & 1.0 & 15.0 & 5.0 & 20.0 & 16.6 & 2.41 &  2.48 & 2.80\\
      ...  &...  &...   &... &...              &...  &...  &...  &...   & ... & ... & ...& ...\\
     m1r1 & 4.3 & $1\times10^{3}$ & $1\times10^{-2}$& 0.04  & 1.0 & 25.0 & 5.0 & 5.0 & 41.7 & 2.61 &  2.64 & 2.54\\
     m1r2 & 4.3 & $1\times10^{3}$ & $1\times10^{-2}$& 0.04  & 1.0 & 50.0 & 5.0 & 5.0 & 52.8 & 2.95 &  3.01 & 3.28 \\
     m1r3 & 4.3 & $1\times10^{3}$ & $1\times10^{-2}$& 0.04  & 1.0 & 75.0 & 5.0 & 5.0 & 39.4 & 3.33 &  3.45 & 3.93\\
     ...  &...  &...   &... &...              &...  &...  &...  &...   & ... & ... & ...& ...\\
     m1w1 & 4.3 & $1\times10^{3}$ & $1\times10^{-2}$& 0.04  & 1.0 & 15.0 & 2.5 & 5.0 & 27.5 & 2.84 &  2.89 & 2.33\\
     m1w2 & 4.3 & $1\times10^{3}$ & $1\times10^{-2}$& 0.04  & 1.0 & 15.0 & 7.5 & 5.0 & 34.8 & 2.43 &  2.50 & 2.36\\
     m1w3 & 4.3 & $1\times10^{3}$ & $1\times10^{-2}$& 0.04  & 1.0 & 15.0 & 10.0 & 5.0 & 32.5 & 2.43 &  2.47 & 2.32\\
     ...  &...  &...   &... &...              &...  &...  &...  &...   & ... & ... & ... &...\\
     m1h1 & 4.3 & $1\times10^{3}$ & $1\times10^{-2}$& 0.03  & 1.0 & 15.0 & 5.0 & 5.0 & 25.9 & 2.31 &  2.37 & 2.39\\
     m1h2 & 4.3 & $1\times10^{3}$ & $1\times10^{-2}$& 0.06  & 1.0 & 15.0 & 5.0 & 5.0 & 35.9 & 2.58 &  2.50 & 2.65\\
     ...  &...  &...   &... &...              &...  &...  &...  &...   & ... & ... & ... &...\\
     m1b1 & 4.3 & $1\times10^{3}$ & $1\times10^{-2}$& 0.04  & 0.5 & 15.0 & 5.0 & 5.0 & 26.4 & 2.10 &  2.12 & 2.37\\
     m1b2 & 4.3 & $1\times10^{3}$ & $1\times10^{-2}$& 0.04  & 1.5 & 15.0 & 5.0 & 5.0 & 45.2 & 2.28 &  2.35 & 2.52\\
     ...  &...  &...   &... &...              &...  &...  &...  &...   & ... & ... & ...&...\\
     nogap & 4.3 & $1\times10^{3}$ & $1\times10^{-2}$& 0.04  & 1.0 & $-$ & $-$ & $-$ & 5.14 & 4.04 &  4.29 & 3.61\\
     \hline\hline
   \end{tabular}
   \end{center}
   \tablecomments{For all models, we adopt typical stellar parameters with $M_{\star}=0.6\ M_{\sun}$, $R_{\star}=2\ R_{\sun}$, $T_{\rm eff}=3850\ {\rm K}$.}
  \end{table*}

\section{Results} \label{sec:results}

In this section, we describe our results for modeling the gas and dust dynamics for 
various disks, and discuss the comparison between the dust continuum observations of 
some young disks observed at mm wavelengths and the predictions of our models. 
Based on the comparisons with global spectral index, disk brightness with current observations, 
we aim at setting some constraints on the disk and coagulation parameter 
(e.g., disk viscosity, fragmentation velocity).  The different behaviors of radial profiles for 
the spectral indices observed at different wavelengths in the future can also be used 
to break the model degeneracy in producing a low spectral index.

We first discuss the dust dynamics based on the coagulation and fragmentation model presented above.
The main model parameters are listed in Table~\ref{tab:para}, otherwise their default values are adopted.

\subsection{Fiducial Model}
Here we start by discussing the model labeled as m1v2 in Table~\ref{tab:para}, which 
is characterized by $\Sigma_{0} = 4.3~\rm{g~cm^{-2}}$ and fragmentation velocity 
$v_{\rm f} = 10^3 ~\rm{cm~s^{-1}}$. The initial disk gas
mass is $0.01\ M_{\odot}$.

After incorporating the low viscosity rectangular gap between 12.5$-$17.5 au as shown 
in Figure~\ref{fig:viscosity}, we ran the numerical simulation for the two-fluid gas and dust 
evolution with dust coagulation and fragmentation for 2.1 Myr, which corresponds to 
14700 orbits at 5 au. The dust settles down into an equilibrium state with a 
balance of coagulation and fragmentation after $\sim2$ Myr.

\begin{figure}[htbp]
\begin{center}
\includegraphics[width=0.45\textwidth]{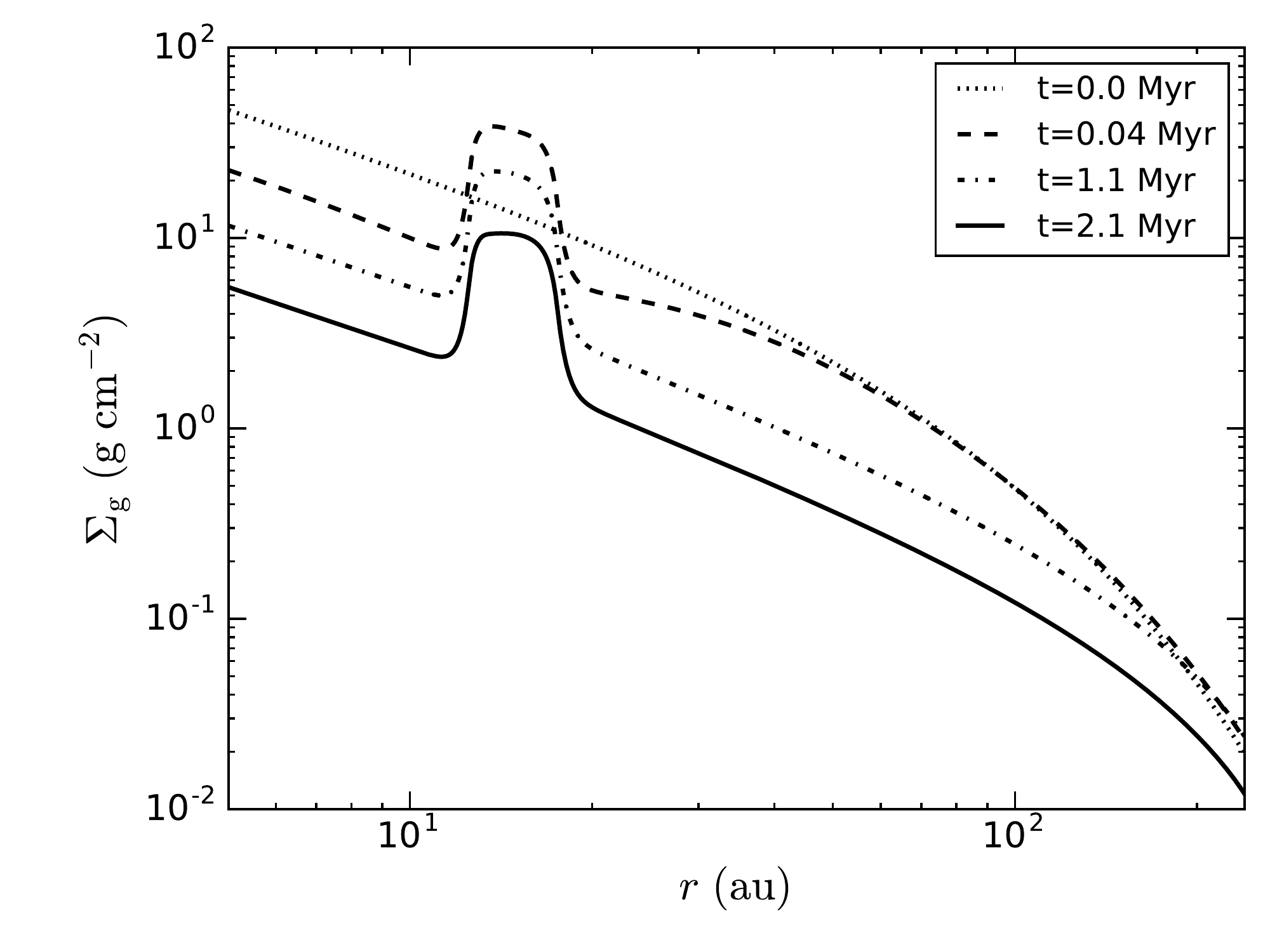}
\hskip 0.5truecm
\includegraphics[width=0.45\textwidth]{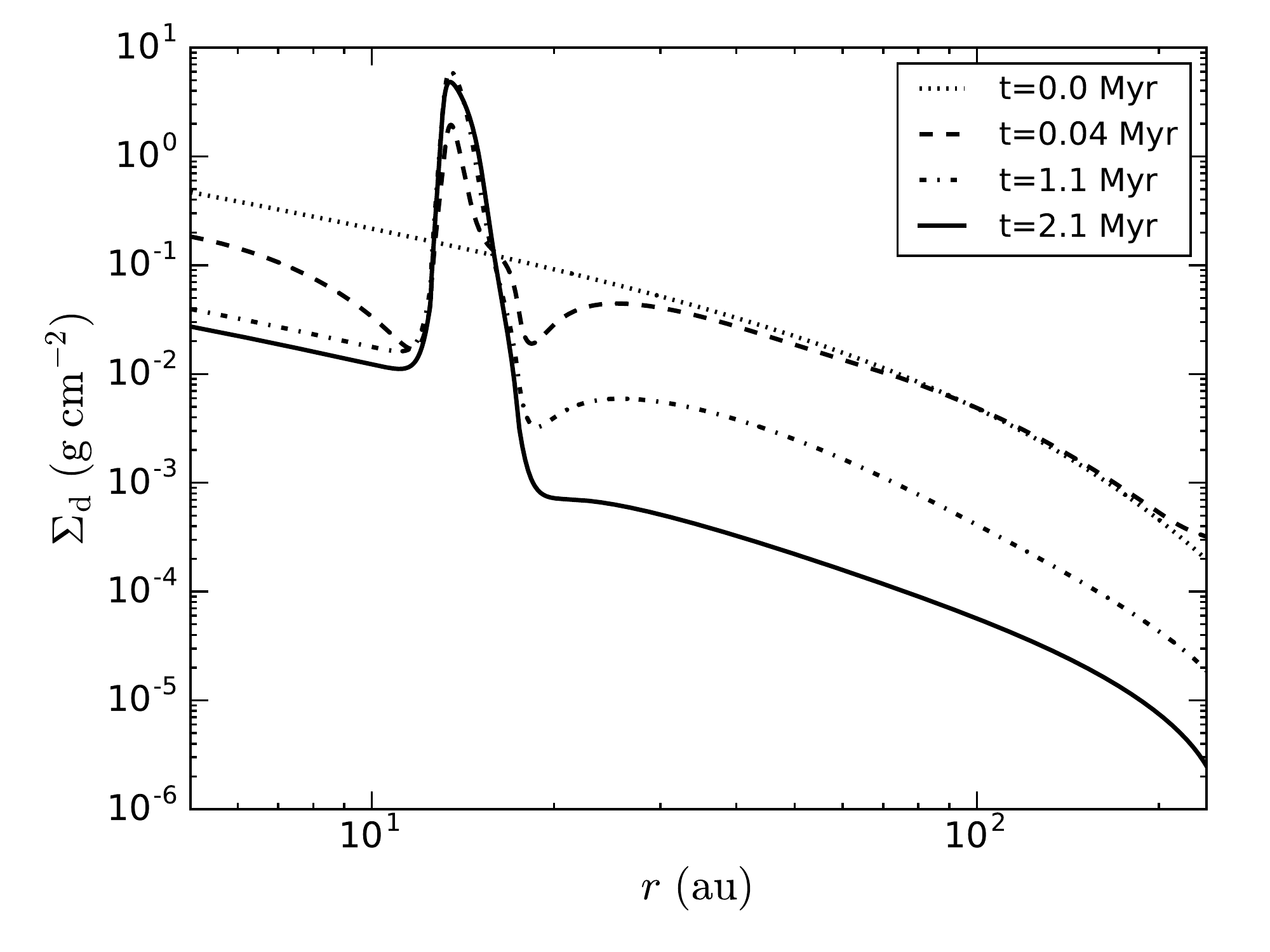}
\hskip 0.5truecm
\includegraphics[width=0.45\textwidth]{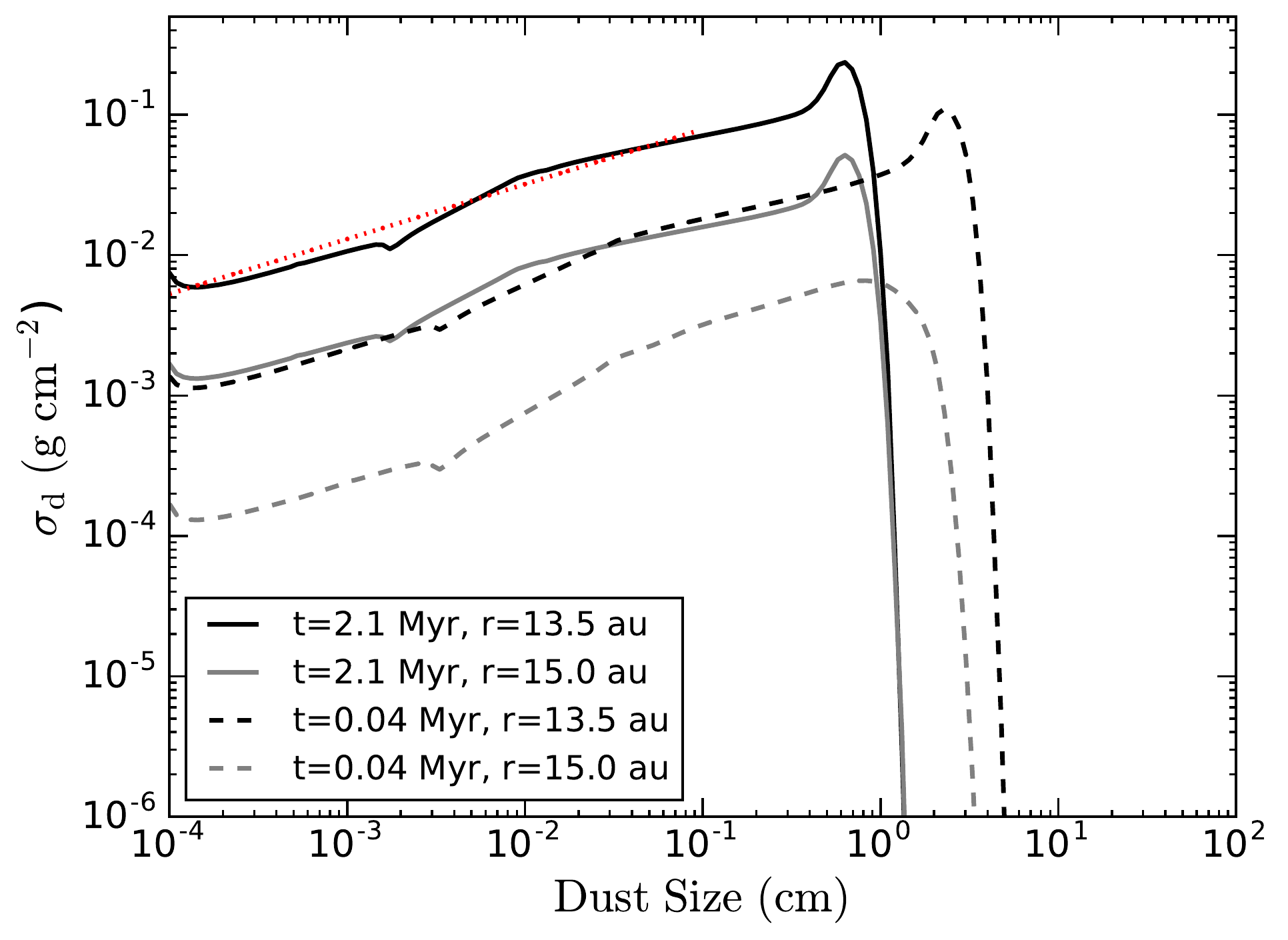}
\end{center}
\caption{Gas and dust radial profiles for the fiducial model m1v2. 
Upper panel: the gas surface density profile at different times. The middle panel 
shows the dust surface density distribution for all dust species as a function of radius and the bottom one 
shows the dust size distribution. The red dotted line in the bottom panel shows the fitted power law with an index of $0.4$ for the size distribution. We adopt a $\alpha_{\rm vis}$ profile with a gap c
entered at 15 au, width $w_{\rm ring}=5\ {\rm au}$ and $d_{\rm ring}=5.0$. 
The gas bump can be established quickly in the region of the viscosity gap. 
Other disk parameters are listed in Table~\ref{tab:para}.
}\label{fig:dyn}
\end{figure}

The radial profile of the gas surface density is shown at different times in the upper 
panel of Figure~\ref{fig:dyn}.  We generically separate this profile into the bump region 
and the smooth region in the following discussions. 
The gas bump is built up quickly in the viscosity gap region.  
In the case of mass conservation without disk wind, steady accretion implies that 
$r\Sigma_{\rm g}v_{r,\rm g}={\rm constant}$, and 
$v_{r,\rm g}\propto\nu_{\rm g}/r=\alpha_{\rm vis} c_{\rm s}h_{\rm g}/r$, 
so that $\Sigma_{\rm g}\propto\alpha_{\rm vis}^{-1}$. Therefore, 
a lower $\alpha_{\rm vis}$ in the gap region leads to a higher gas surface density. The minimum Toomre Q parameter ($\sim10$) is reached in the gas bump region at the initial stage, justifying that the disk is gravitational stable during the whole evolution.

\begin{figure}[htbp]
\begin{center}
\includegraphics[width=0.45\textwidth]{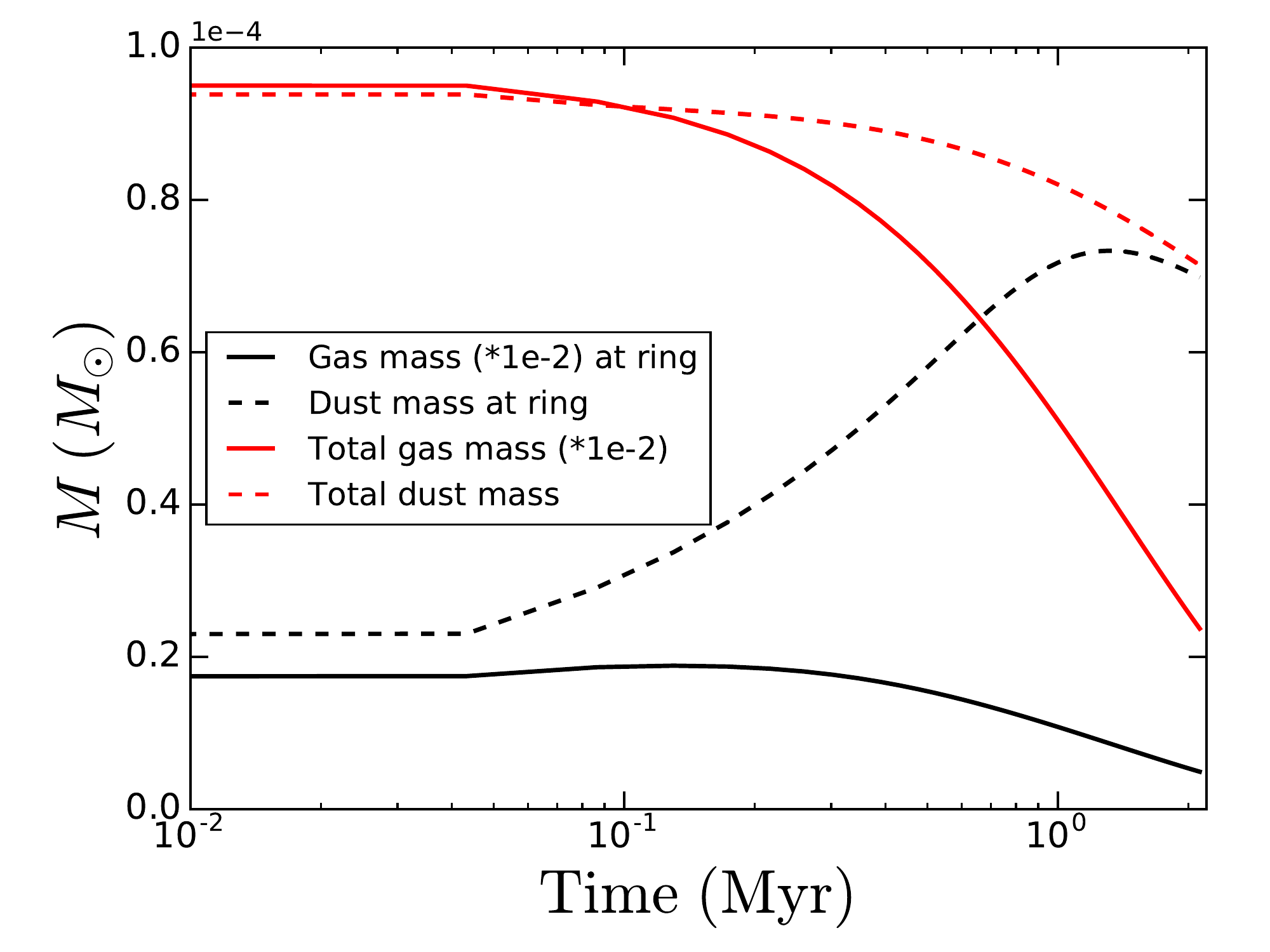}
\end{center}
\caption{The time evolution of the disk gas and dust mass. 
Solid (dashed) lines correspond to  gas (dust) mass, respectively. 
Black  lines show the mass from the ring region, while red ones show the total disk mass. 
For the purpose of presentation, the gas masses are multiplied by $10^{-2}$.
}\label{fig:evo}
\end{figure}

\begin{figure}[htbp]
\begin{center}
\includegraphics[width=0.45\textwidth]{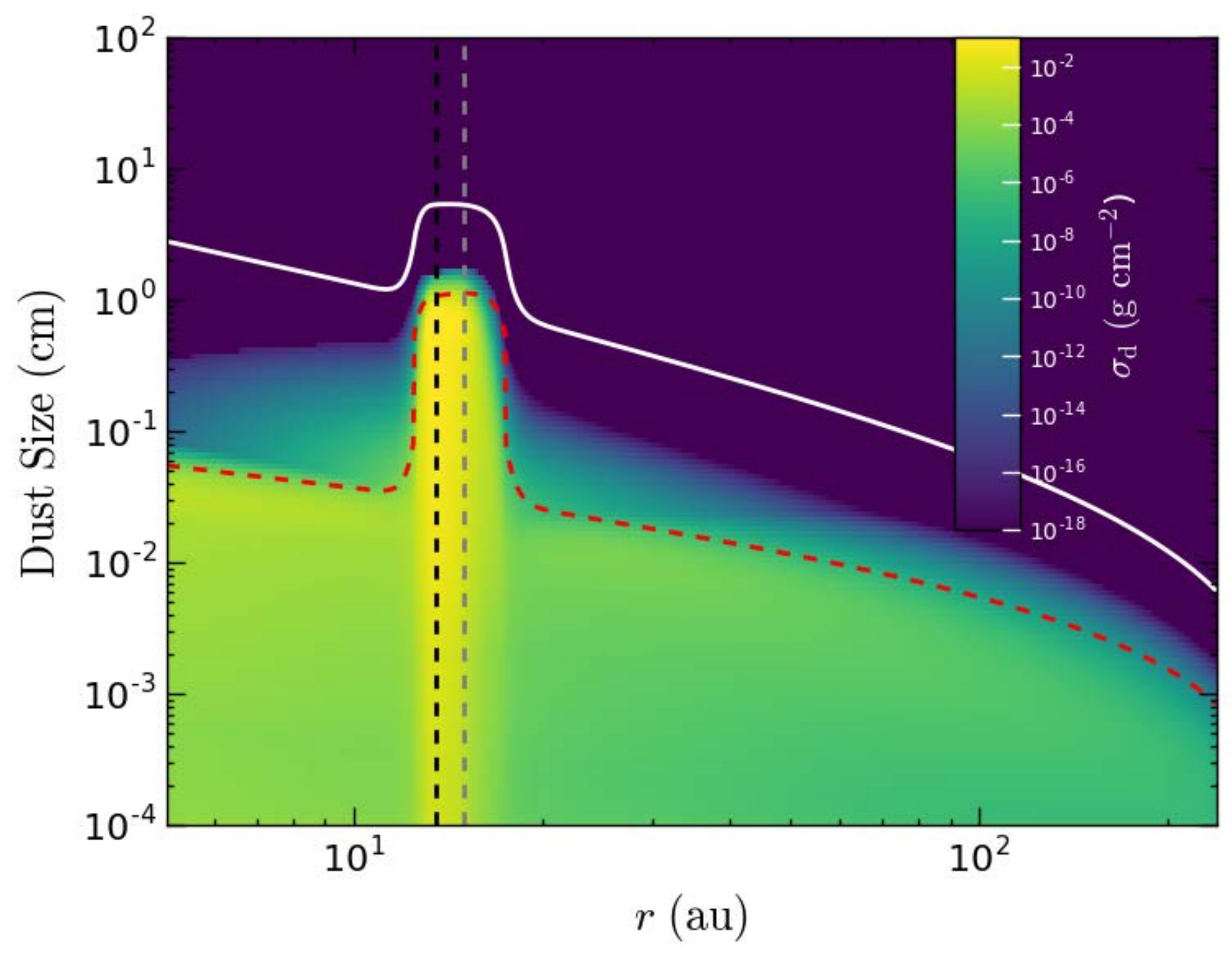}
\end{center}
\caption{Dust surface density distribution for different dust species at the time 2.1 Myr for model m1v2. 
The red dashed line indicates the fragmentation barrier, which represents the maximum size 
of the particles before they reach fragmentation velocities.  The white solid line denotes 
the grain size corresponding to a Stokes number of unity, which shows the same shape 
as the gas surface density $\Sigma_{\rm g}(r)$. The black and grey vertical dashed lines 
indicate the location of $r=13.5\ {\rm au}$ and $r=15.0\ {\rm au}$, respectively, 
between which we extract the dust size distributions plotted
in the lower panel of Figure~\ref{fig:dyn}.}\label{fig:surf_dust}
\end{figure}

\begin{figure}[htbp]
\begin{center}
\includegraphics[width=0.45\textwidth]{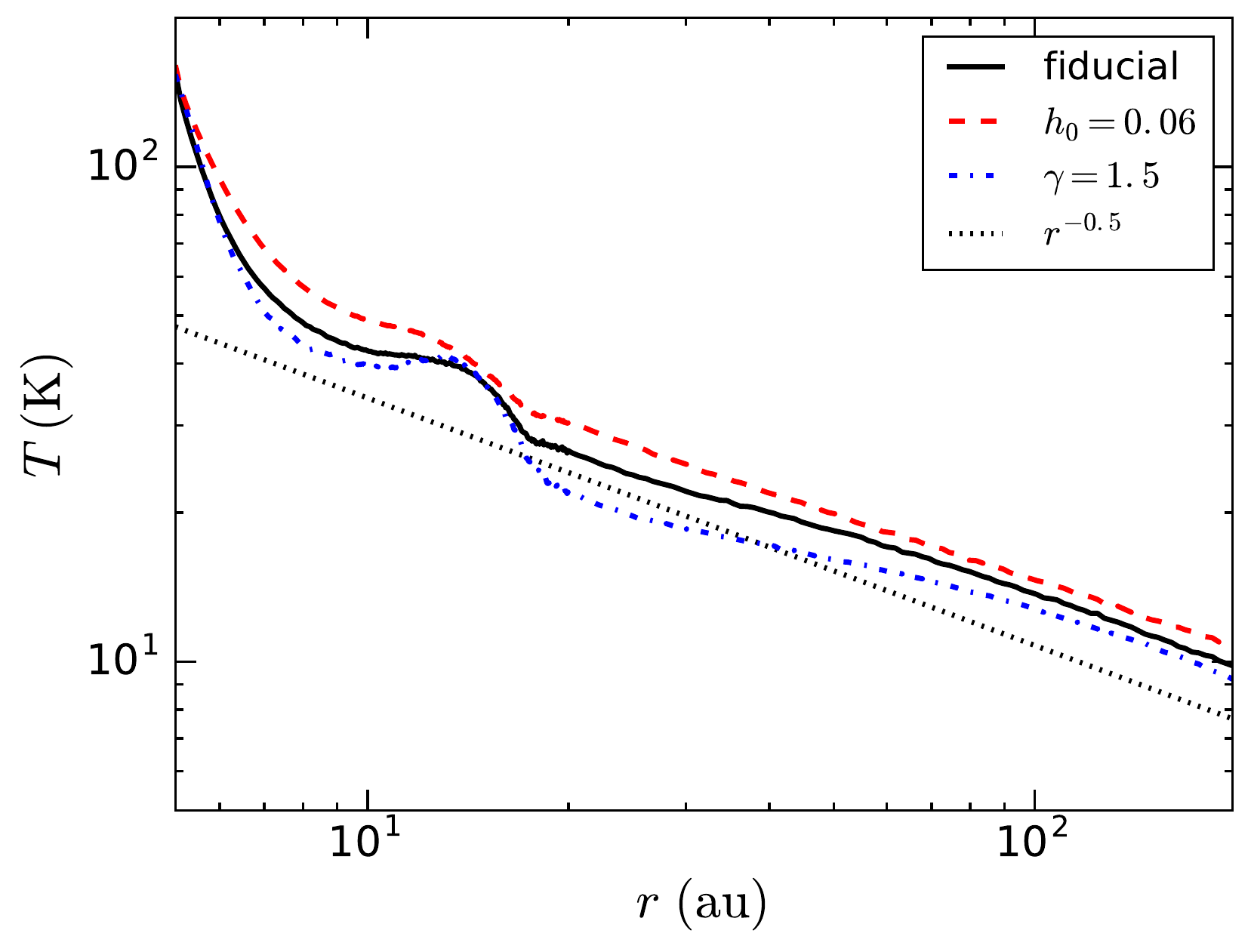}
\end{center}
\caption{Radial profile of the dust temperature at the equatorial plane of disk for our fiducial model. For a comparison, we also show the temperature profiles with different disk scale height $h_{0}$ and gas surface density profile index $\gamma$. The dust temperature profile in the outer disk is close to the $r^{-0.5}$ power law shown as the black dotted line, which is the temperature profile of our fiducial model adopted in hydrodynamical simulations based on Equation~(\ref{eq:cs}). }\label{fig:temp}
\end{figure}

We also show the radial distribution of the total dust surface density $\Sigma_{\rm d}$, 
which sums up all the dust surface densities of each species $\sigma_{\rm d}$,   
in the middle panel of Figure~\ref{fig:dyn}. The dust accumulation in the gas bump 
region is clearly shown in the radial distribution of the dust surface density plot. 
We can see that the dust-to-gas ratio can increase from the initial value of 0.01 to 0.5 
at the final stage in the dust trapping region, where the dust feedback could become important \citep{Fu2014,Miranda2017}.

A well-defined power-law distribution for the sizes of the dust particles in the 
dust trapping region is shown in the lower panel of Figure~\ref{fig:dyn}. 
The red dotted line shows the fitted power law with an index of $0.4$, 
which is equivalent to a power-law index of $q=3.6$  for the number density distribution 
$n(a)\propto a^{-q}$, close to the value of 3.5 for the dust size 
distribution in the ISM \citep{Mathis1977}.

The evolution of the total disk gas and dust masses is shown in Figure \ref{fig:evo}, where
the masses within the ``bump" region 12.5$-$17.5 au are plotted as well. 
As expected, the total disk mass has decreased significantly due to accretion during
the 2.1 Myr evolution. The relative fraction of disk gas mass within the bump has increased.
The total dust mass has reduced by about $30\%$ and, by 2.1 Myr, most of the 
dust mass is now residing within the bump region. The ring is essential to retain the
dust mass from the rapid radial drift and loss through the disk boundary. 

We further show the 2D contour plot of the dust surface density distribution for all 
dust species at all stellocentric radii throughout the disk in Figure~\ref{fig:surf_dust}. 
The fragmentation barrier, which was defined in Equation~(\ref{eq:amax_frag}), 
is represented as the red dashed line. It can be seen that the bump in the fragmentation 
barrier that limits the dust size growth can be attributed to both the smaller viscosity 
and the larger gas surface density.
The grain sizes corresponding to $a_{\rm St=1}$ (ref. to Equation~(\ref{eq:amax_st})) 
are shown as the white line, which has the same slope as the gas surface density, 
i.e. a ringed structure around 15 au as in the upper panel of Figure~\ref{fig:dyn}. 
The fact that all the grain species in this model lie below the white line representing 
$a_{\rm St=1}$ indicates that ${\rm St} <1$ applies to all grains.

As expected, the dust spatial distribution shows significant trapping in the 
gas bump region as shown in Figure~\ref{fig:surf_dust} (see also the middle 
panel of Figure~\ref{fig:dyn}). This is because the radial inward motion of the dust is 
significantly slowed down as a result of the positive pressure $(\Sigma_{\rm g}c_{\rm s}^{2})$ 
gradient in the inner edge of the gas bump. The large local concentration of dust grains 
retained in the gas bump can also facilitate the growth of dust particles from the 
initial 1 $\mu$m size to a few cm due to frequent collisions and sticking of small grains.
Some particles above the fragmentation barrier shown in Figure~\ref{fig:surf_dust} is due to particle diffusion. The larger particles drift faster, which leads to a sharper trap in the inner edge of the bump for the larger grains.
Both dust trapping and size growth are important in determining the dust continuum emission.

Outside the gas bump region, the dust surface density decreases significantly with 
time due to its radial drift. In addition, the fragmentation barrier defined as in 
Equation~(\ref{eq:amax_frag}) is significantly lower outside the bump region than inside the 
bump due to the radial viscosity profile we adopt, which leads to the average dust size 
(weighted by the dust surface density) being much smaller in the smooth region of the disk.
This also explains an increase of dust size with a decreasing stellocentric radius $r$ 
because the dust surface density becomes increasingly higher towards the inner region of 
the disk which results in a higher fragmentation barrier.

We then calculate the dust temperature using \texttt{RADMC-3D} as 
described in Section~\ref{sec:radmc}. The radial profile of the temperature in 
the mid-plane of the disk is shown in Figure~\ref{fig:temp}. 
The temperature profile is well consistent with the $r^{-0.5}$ scaling expected 
for a disk heated by the central star except for a bump around 15 au \citep{Armitage2010}. 
The dust temperature profile is also roughly in agreement with the profile in our 
hydrodynamical simulation based on Equation~({\ref{eq:cs}}), which is shown as 
black dotted line in Figure~\ref{fig:temp}.
The temperature bump is  associated with the dust bump feature in the same region, 
which results from the more efficient absorption \citep{Bjorkman2001}. 
A higher dust temperature in the bump could decrease the dust size in that region. 
But this is a small modification in dust size so it cannot significantly influence the 
dust emission, since the current maximum size is much bigger than $\sim1{\rm mm}$. 
It is expected that the temperature in the mid-plane is lower than that at the disk surface 
in the outer disk region, while the temperature in the inner edge of the disk is close to 
the temperature at the disk surface determined by the stellar heating. 
This results in the steep slope for the mid-plane temperature profile in the inner disk. 
The total emission from the dust in this very inner region is, however, negligible due to 
the significant dust depletion as discussed above.

\begin{figure}[htbp]
\begin{center}
\includegraphics[width=0.45\textwidth]{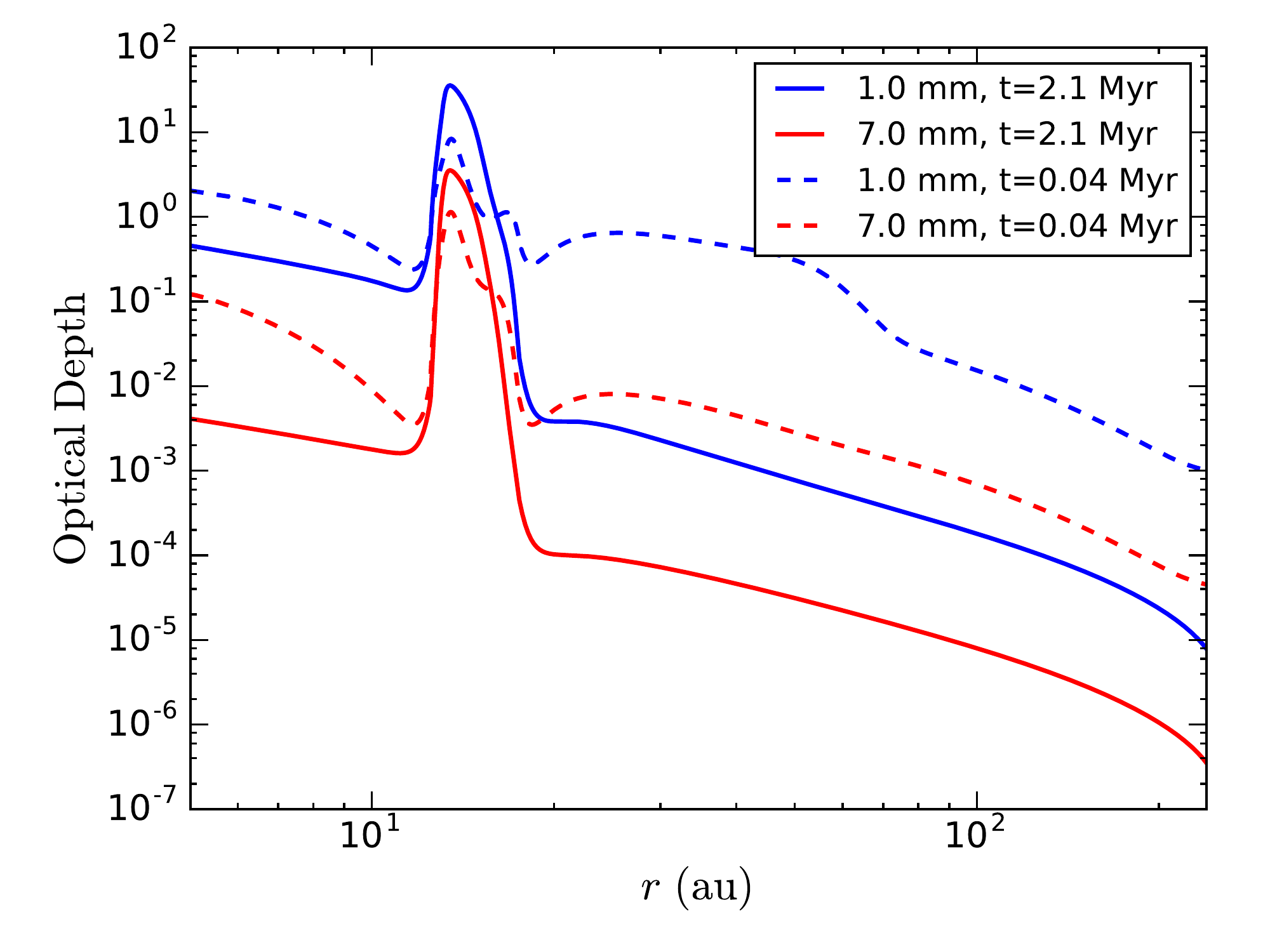}
\includegraphics[width=0.45\textwidth]{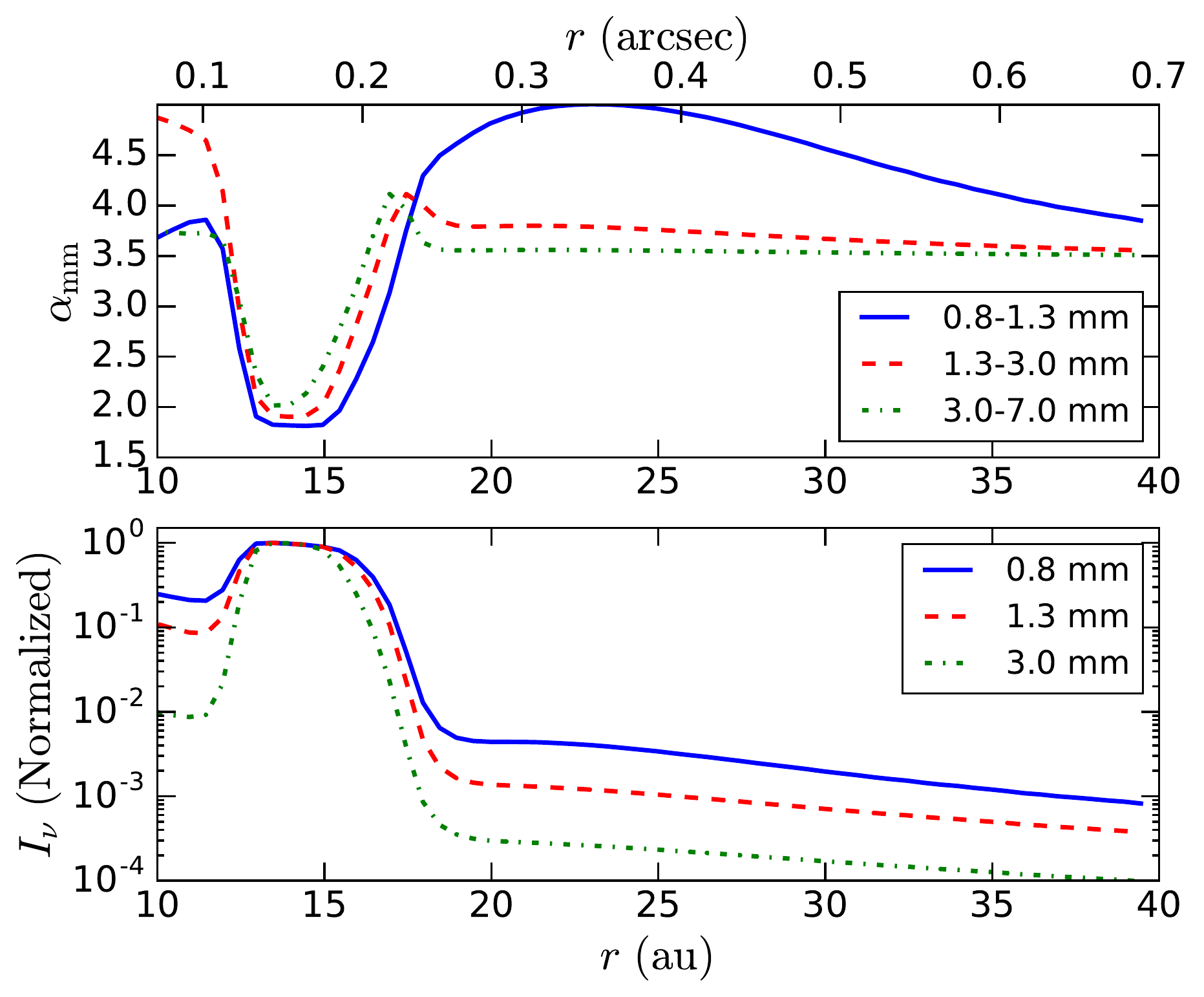}
\end{center}
\caption{Radial profiles of optical depth (upper panel), spectral index, and the normalized flux (lower panel) at three different wavelengths for model m1v2. Two evolution stages (dashed: 0.04 Myr; solid: 2.1 Myr) are chosen to show the time evolution of the optical depth at two wavelengths (blue: 1.0 mm; red: 7.0 mm). The spectral indices are calculated at three different wavebands at 2.1 Myr, represented as different lines. }\label{fig:optical_depth}
\end{figure}

Due to the dust accumulation in the gas bump region, the local effective optical depth 
$\tau_{\rm eff}$ increases significantly. The radial profiles of the effective optical depth 
$\tau_{\rm eff}$ for all dust species, which are obtained by summing the optical depth 
over all dust species, are shown in the upper panel of Figure~\ref{fig:optical_depth}. 
We show the optical depth at two wavelengths, i.e., 1 and 7 mm, and at two epochs, close to 
the initial stage (0.04 Myr) and at the end of the simulation (2.1 Myr).
At the wavelength of 1 mm, the dust-trapping region is optically thick, while other regions 
are still optically thin at the end of the simulation (2.1 Myr).  At the longer wavelength of 7 mm, 
the emission becomes marginally optically thick even in the dust trapping region. Note that the optical depth in the ring is significant higher than that inferred from DSHARP samples \citep{Dullemond2018}. A lower gas surface $\Sigma_{0}$ can make a better agreement with these observations. Furthermore, the inclusion of dust scattering can result in an optically thick disk appearing as optically thin \citep{Zhu2019}.
The dust optical depth in the trapping region does not evolve significantly with time, 
however, it decreases significantly outside the trapping region at mm wavelengths. 
This is mostly due to the dust radial drift effect.

\begin{figure}[htbp]
\begin{center}
\includegraphics[width=0.45\textwidth]{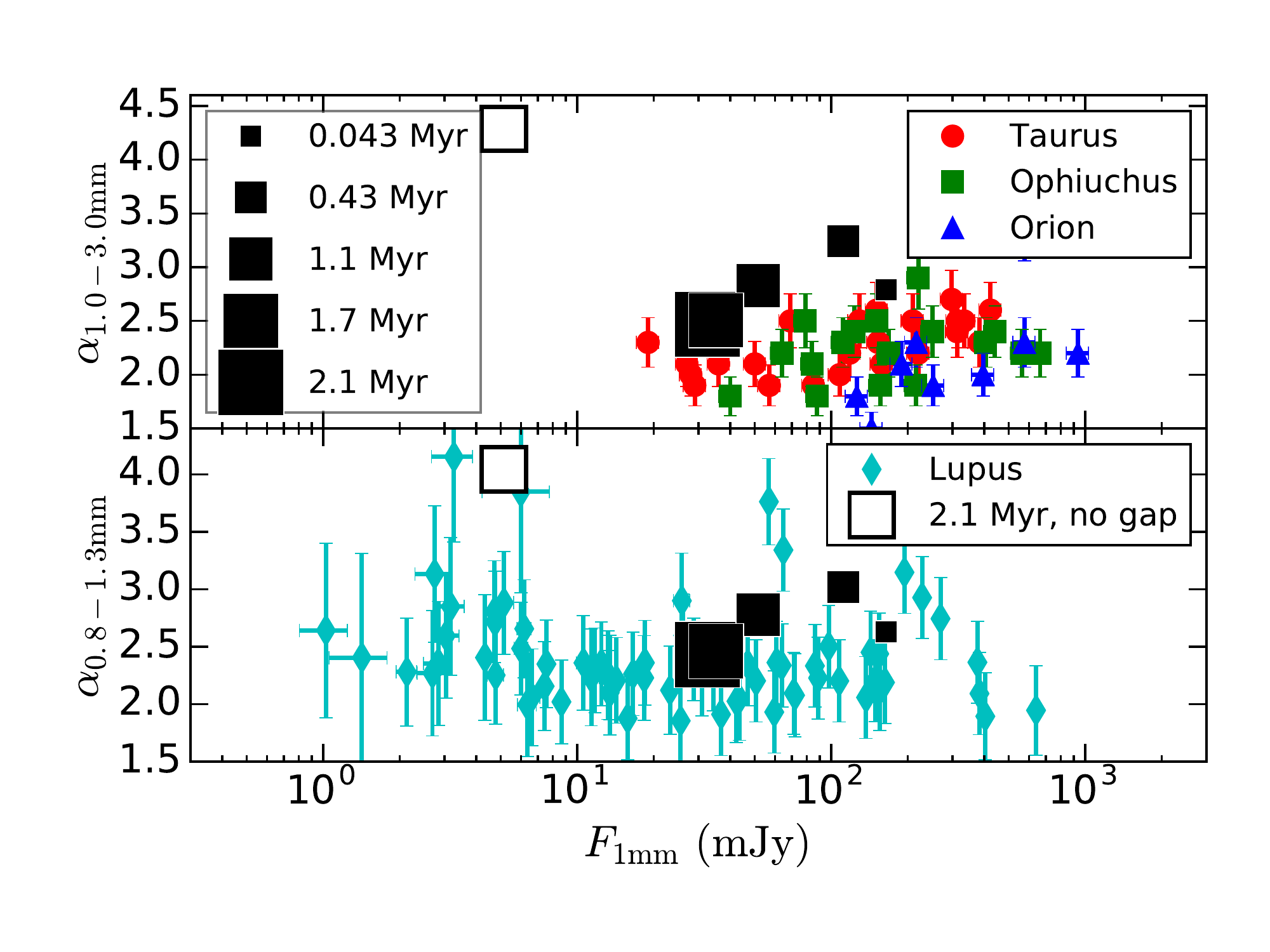}
\end{center}
\caption{Comparison between the time evolution of 1 mm flux $F_{\rm 1mm}-\alpha_{\rm mm}$ scatter plot from our model m1v2 with the observations from four star-forming regions \citep{Ricci2012,Ansdell2018}.  The model without any viscosity gap (model nogap) is represented as open square. The observed fluxes are all normalized with a disk distance of 140 pc and extrapolated to 1 mm based on the observed spectral indices. The spectral indices are calculated from the flux between 1.0 mm and 3.0 mm (upper panel), and between 0.89 mm and 1.33 mm (lower panel).  Some other key model parameters are listed in Table~\ref{tab:para}, otherwise the default values are adopted. }\label{fig:fa_m1v2}
\end{figure}

After we obtained the dust size and radial distribution, its temperature, and optical depth for 
all dust species, we can obtain the radial profile of the dust continuum emission at different sub-mm and mm bands with a detailed radiative transfer calculation using \texttt{RADMC-3D}, which is shown in the lower panel of Figure~\ref{fig:optical_depth}.
The flux at different bands can be obtained by integrating the dust continuum emission 
over the whole disk. We can then calculate the corresponding global spectral index and 
the 1mm dust continuum flux. The results for model m1v2 are shown in Table~\ref{tab:para} 
and Figure~\ref{fig:fa_m1v2}. In Figure~\ref{fig:fa_m1v2}, we show the time variation of the 
spectral index calculated between three different sub-mm/mm wavelengths as a function 
of the 1 mm flux density. The 1 mm flux decreases significantly at the early stage 
of the evolution, and then show less temporal variations after $\sim$1.7 Myr. 
Such a decreasing flux is a consequence of the dust radial evolution as shown 
in the middle panel of Figure~\ref{fig:dyn}. In the early stages of the disk evolution, 
the region outside the trapping region contributes a comparable fraction to the total 
flux of the disk. The outer disk contribution is significantly reduced 
as time evolves to $\sim$1.7 Myr. This is also confirmed by the radial profile of the optical 
depth as shown in the upper panel of Figure~\ref{fig:optical_depth}.   
The emission outside the bump region could be initially optically thick for a 
massive disk (as is the case for model m1v2, see the upper panel of Figure~\ref{fig:optical_depth}), 
and it then becomes optically thin due to the dust radial drift. 
The spectral slope from the optically thin emission in the smooth region can be as large as 
about 3.5 since its value is determined by a population of small-sized dust, as shown in the lower panel of Figure~\ref{fig:optical_depth}.  
Therefore, the global spectral index could stay at a low value initially due to 
a large fraction of optically thick emission. It then increases with time rapidly when 
the emission begins to transfer to the optically thin regime, as long as the flux 
contribution from the smooth region is still comparable to that from the bump.   
As disk evolves further, the contribution from the smooth region becomes increasingly negligible, 
which means that the global spectral index is mainly determined by the emission from the bump, 
then the global spectral index decreases with a decreasing flux as shown in 
Figure~\ref{fig:fa_m1v2}.

In addition to the spectral index $\alpha_{\rm 1.0-3.0mm}$, we further calculate the 
index between $0.89-1.3$ mm, i.e., $\alpha_{\rm 0.89-1.3mm}$, which can be compared 
with the observations from the ALMA Lupus survey \citep{Ansdell2018}. 
It can be seen that the time evolution patterns of the spectral index in these two spectral 
windows are quite similar due to the closeness of the two windows.

Note that the opacity we have adopted \citep{Isella2009,Ricci2010a} is higher by a 
factor of a few ($\sim5$) than that adopted in other works \citep[e.g.,][]{Semenov2003,Birnstiel2018}. 
To test the sensitivity of our spectral results on the choice of dust opacity,
we adopt a norm silicates (NRM) dust opacity model\footnote{http://www2.mpia-hd.mpg.de/homes/henning/Dust\_opacities/Opacities\\/RI/new\_ri.html}. It turns out that the global spectral index around 
1 mm only changes only slightly and the final disk flux $F_{\rm 1mm}$ decreases by 
a factor of $\sim2$. The decrease of the total flux is mainly because the optical depth in the 
trapping region becomes lower. Such modifications do not significantly change our main conclusions. 
In the following discussion, we will only adopt the opacity from \citeauthor{Isella2009} (\citeyear{Isella2009}; \citealt{Ricci2010a}) unless otherwise stated.

\subsection{Dependence on Parameters}

In this section, we study how our main results depend on the model parameters we have used. 
The main parameters we will explore include the fragmentation velocity $v_{\rm f}$, 
the initial gas surface density $\Sigma_{0}$, disk global viscosity $\alpha_{0}$, 
viscosity gap (ring) width $w_{\rm ring}$, gap (ring) depth $d_{\rm ring}$, 
gap (ring) location $r_{\rm ring}$, disk scale height $h_{0}$, and disk gas profile index $\gamma$, 
which are listed in Table~\ref{tab:para}. In the following, we evolve all the disk models for 
2.1 Myr when the dust/gas dynamics has reached the equilibrium state.

\subsubsection{Fragmentation Velocity}

\begin{figure*}[htbp]
\begin{center}
\includegraphics[width=0.45\textwidth]{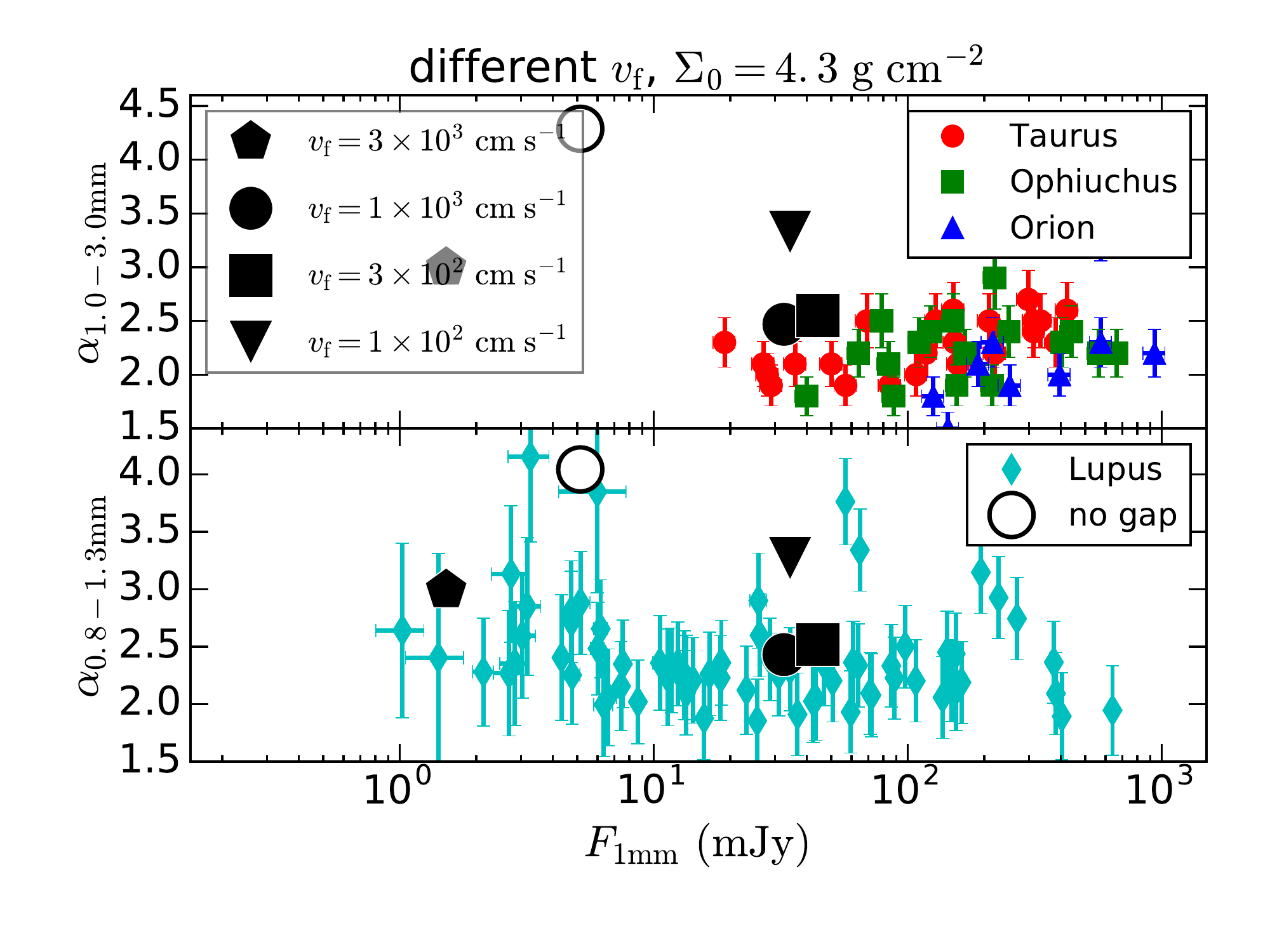}
\includegraphics[width=0.45\textwidth]{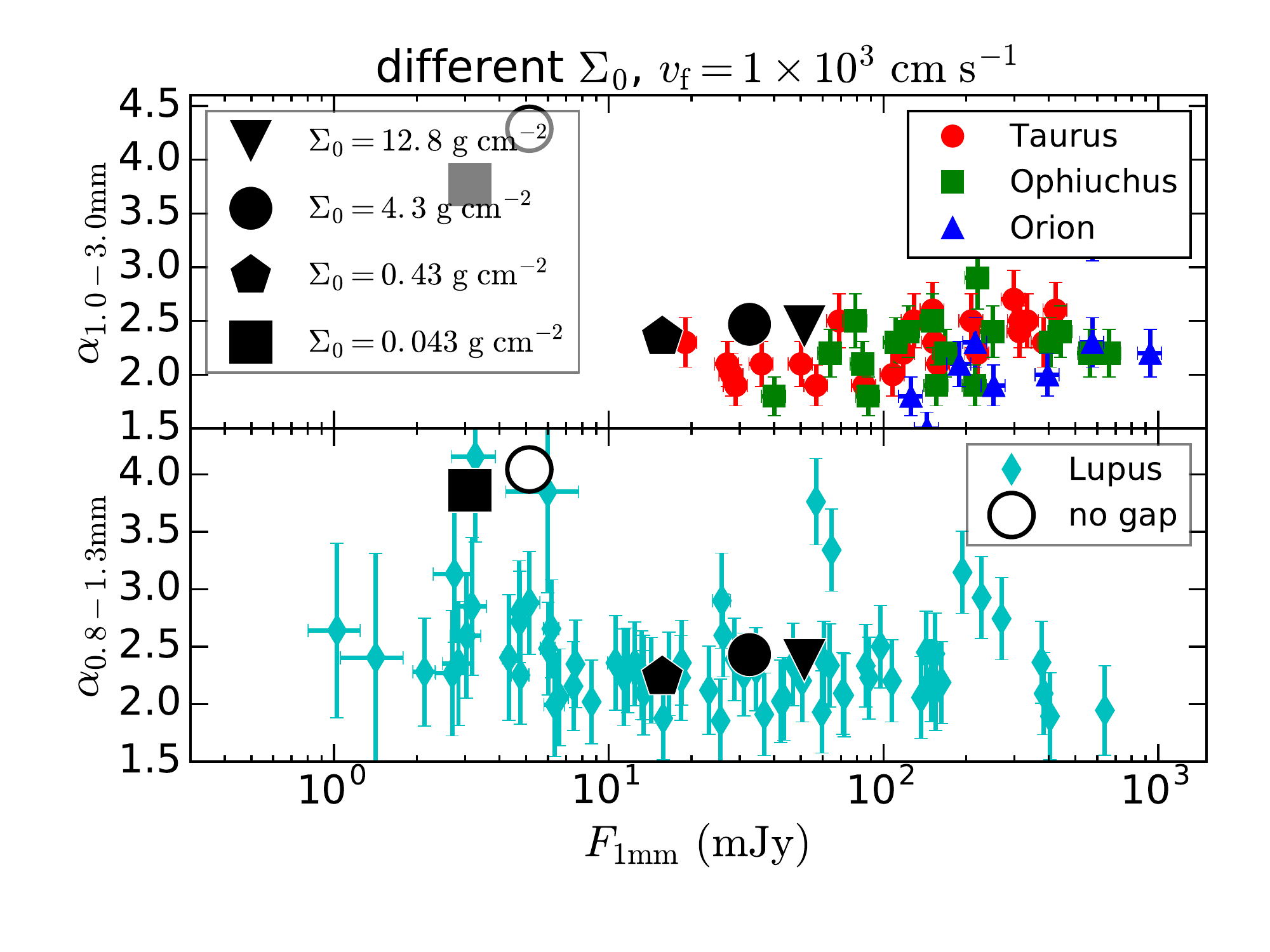}
\includegraphics[width=0.45\textwidth]{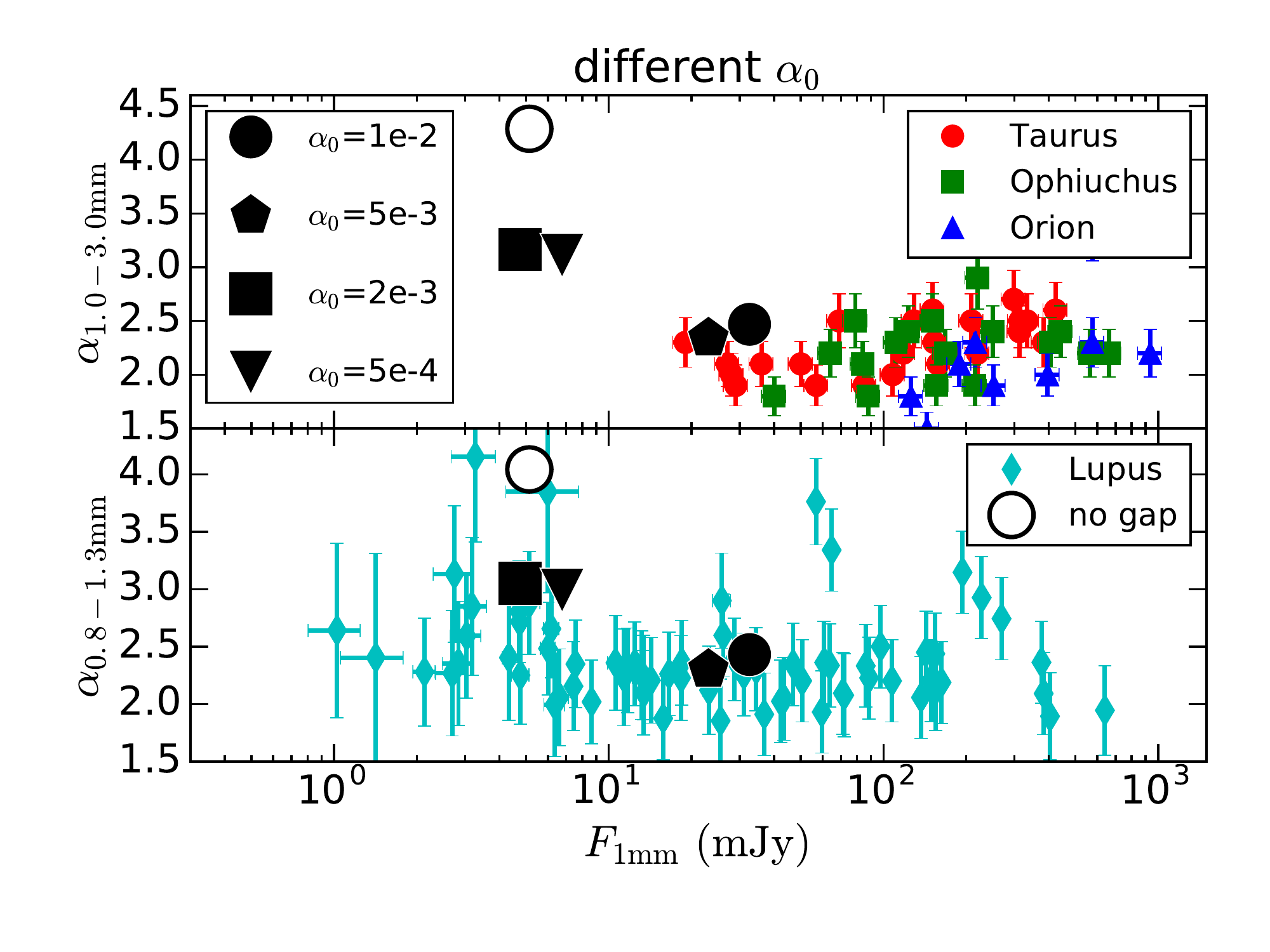}
\includegraphics[width=0.45\textwidth]{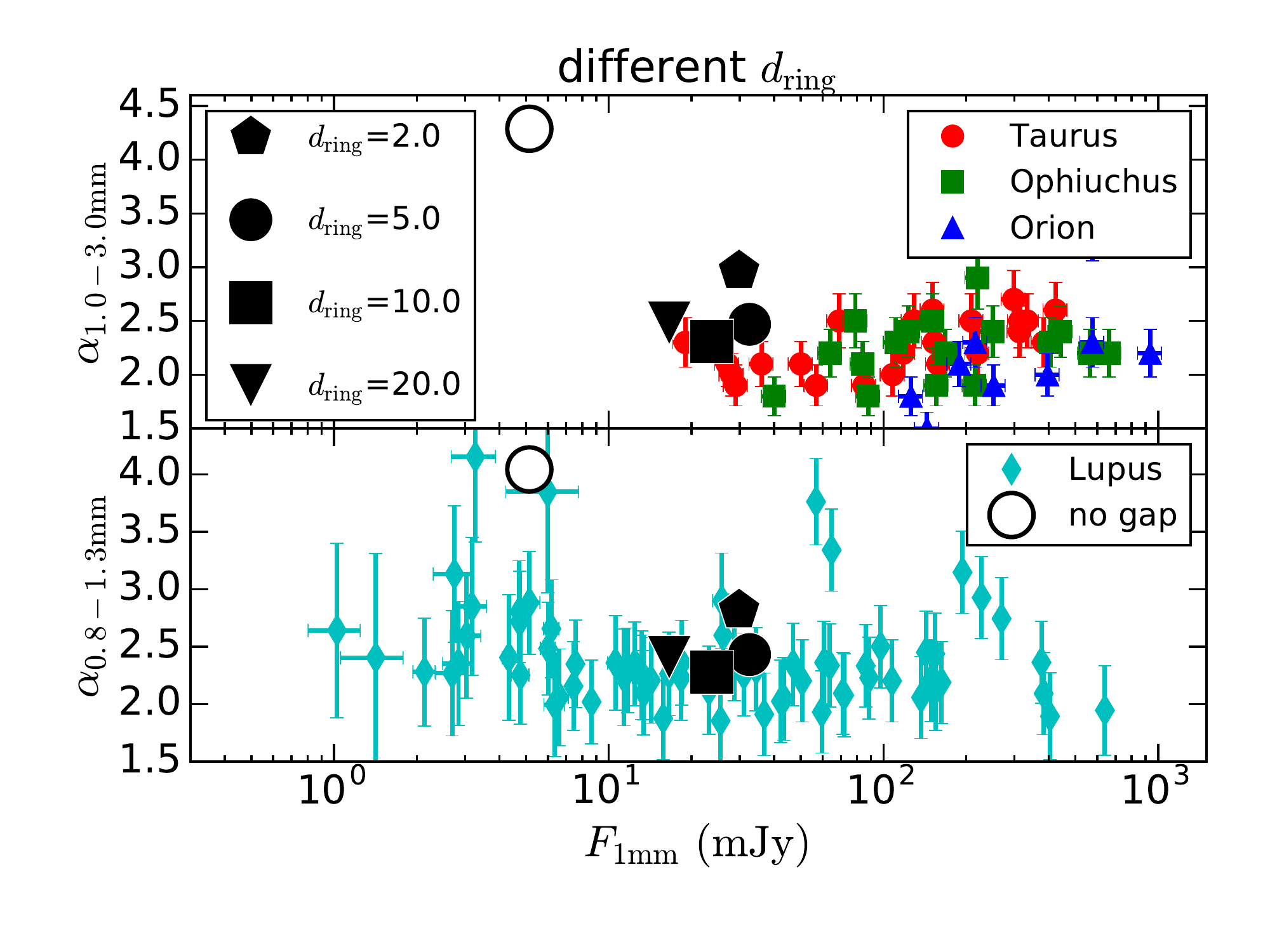}
\includegraphics[width=0.45\textwidth]{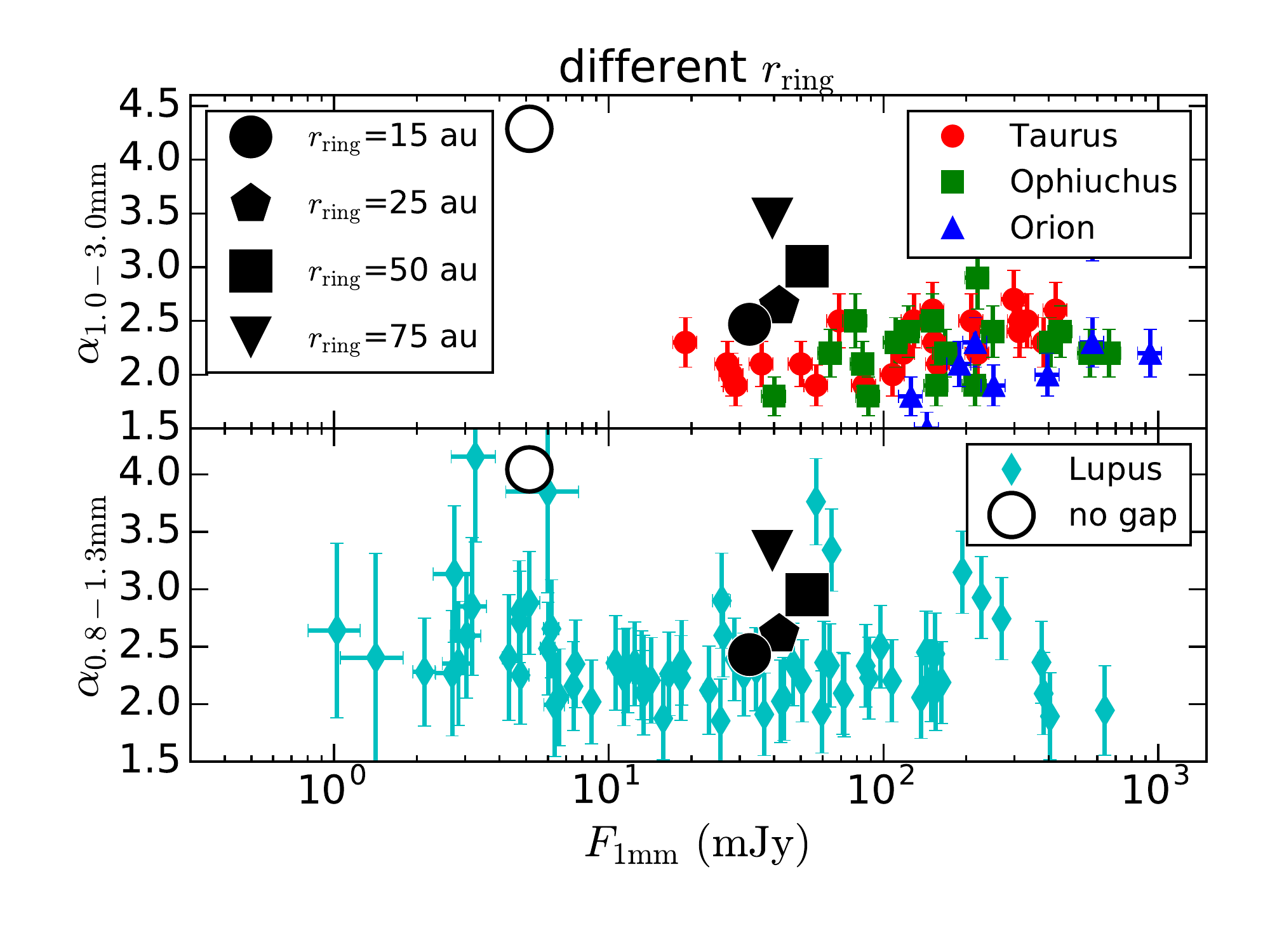}
\includegraphics[width=0.45\textwidth]{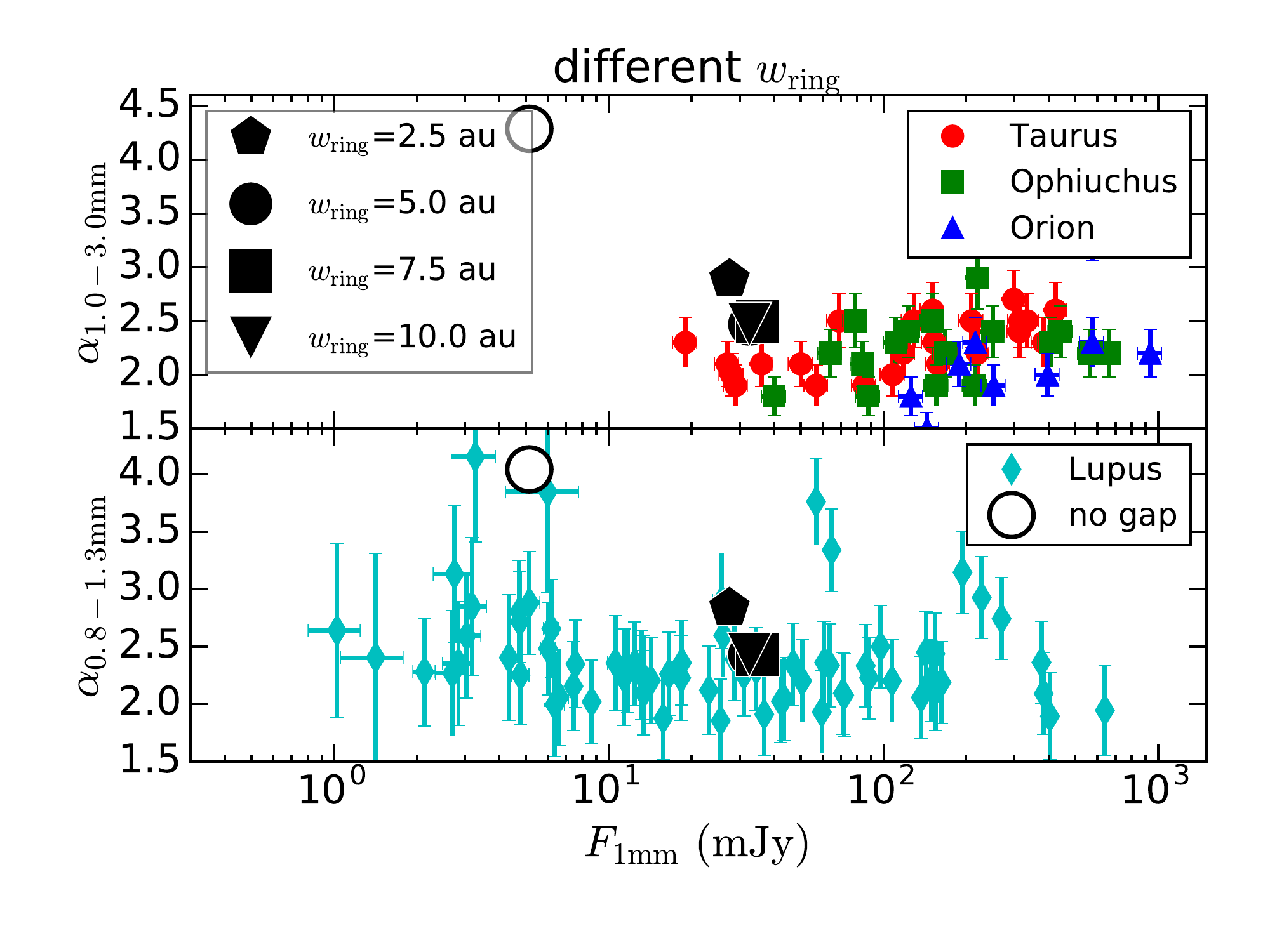}
\end{center}
\caption{The model parameter dependence of the 1 mm continuum flux $F_{\rm 1mm}$ and the mm spectral index $\alpha_{\rm mm}$. The spectral index at two wavebands, $\alpha_{\rm 1.0-3.0mm}$ and $\alpha_{\rm 0.8-1.3mm}$, are calculated for each model.  The observations data are also collected from four star-forming regions \citep{Ricci2012,Ansdell2018}, same as Figure~\ref{fig:fa_m1v2}.  The six panels correspond to models with different fragmentation velocity $v_{\rm f}$ (upper left), disk mass $\Sigma_{0}$ (upper right), disk global viscosity $\alpha_{0}$ (middle left), viscosity gap depth $d_{\rm ring}$ (middle right), viscosity gap location $r_{\rm ring}$ (bottom left), and viscosity gap width $w_{\rm ring}$ (bottom right), respectively. The model without any viscosity gap in each panel (model nogap) is represented as open circle. Some other key model parameters are also listed in Table~\ref{tab:para}, otherwise their default values are adopted. Note that a common feature  
is that the models do not explain the bright disks. }\label{fig:flux_alpha}
\end{figure*}

We first study the effect of fragmentation velocity $v_{\rm f}$ of the dust coagulation model.
We ran several models with $v_{\rm f}$ in the range of $1\times10^{2}$ to
$3\times10^{3}\ {\rm cm\ s^{-1}}$ while fixing all other parameters.
The final global spectral index and 1 mm flux for models with varying fragmentation 
velocity are shown in the upper left panel of Figure~\ref{fig:flux_alpha}.

In comparison with the case of $v_{\rm f}=1\times10^{3}\ {\rm cm\ s^{-1}}$ in Figure~\ref{fig:fa_m1v2},  
the spectral index around 1 mm $\alpha_{\rm 1.0-3.0mm}$ is insensitive to $v_{\rm f}$ 
until  $v_{\rm f}$ decreases to $1\times10^{2}\ {\rm cm\ s^{-1}}$. 
The spectral index $\alpha_{\rm 1.0-3.0mm}$ for a very small 
$v_{\rm f}=1\times10^{2}\ {\rm cm\ s^{-1}}$ can reach a large value of $\sim3.5$.
The two effects contribute to this large index. As $v_{\rm f}$ becomes very small (i.e., lowering the
fragmentation barrier), it limits the maximum grain size obtained from coagulation. 
With the decrease of $v_{\rm f}$ by one order of magnitude, the maximum dust size decreases 
by two orders of magnitude, giving the largest grain sizes around only $\sim2$ mm. 
In addition, without the large dust grains, the dust emission contributed by the small dust 
becomes optically thin.  Such a large population of small-sized grains with optically thin 
emission thus produces emission with a high spectral index.

When the fragmentation velocity becomes even larger ($v_{\rm f}=3\times10^{3}\ {\rm cm\ s^{-1}}$), 
most of the dust mass now resides in the largest dust grains, reducing the overall surface density
in smaller grains. Because opacity is typically dominated by the small grains, 
the effective optical depth can get significantly reduced, i.e., becoming mostly optically thin. 
As a result, the total mm flux, which is dominated by the dust bump region, 
also decreases significantly when $v_{\rm f}$ becomes large enough, as shown in the upper
left panel of Figure~\ref{fig:flux_alpha}.

The spectral index $\alpha_{\rm 0.8-1.3mm}$ for different $v_{\rm f}$ is also shown in 
Figure~\ref{fig:flux_alpha}, which is similar to $\alpha_{\rm 1.0-3.0mm}$.
The sub-mm/mm fluxes for most of the faintest disks in the Lupus sample 
can be reproduced by models with the high fragmentation velocity $\sim30\ {\rm m\ s^{-1}}$. 
However, it is difficult to use models with high fragmentation velocity  
to explain the disks with low spectral indices as discussed below.

\subsubsection{Disk Mass}

Another important disk parameter is the disk mass, which, in our models, is determined 
by the gas surface density $\Sigma_{0}$. 
In addition to the fiducial model, we chose three other values for $\Sigma_{0}$ 
as listed in Table~\ref{tab:para}. The most massive disk corresponding to 
$\Sigma_{0}=12.8\ {\rm g\ cm^{-2}}$ (model labels including m0) has a total disk mass of 
$0.03\ M_{\odot}$. We also find that, for this model, the minimum Toomre Q parameter in the 
gas bump region can approach 3 during its evolution. 
The dependence of the total disk flux $F_{\rm 1mm}$ on $\Sigma_{0}$ is rather straightforward. 
An increasing $\Sigma_{0}$ will result in a higher $F_{\rm 1mm}$ as shown in the upper 
right panel of Figure~\ref{fig:flux_alpha} because more dust can be available initially.  
But the increasing flux is not linearly related with $\Sigma_{0}$, 
especially when the emission is dominated by the optically thick regions. 
We find that the disk flux is very likely saturated at a flux level of $\sim100$ mJy, 
as shown in Table~\ref{tab:para} (and also the upper right panel of Figure~\ref{fig:flux_alpha}). 
In this optically thick regime, the global spectral index is insensitive to $\Sigma_{0}$ since 
this is determined solely by the spectral slope of the Planck function, and does not 
depend on the size of the emitting dust.

When the initial dust surface density is decreased by two orders of magnitude 
through the decrease of $\Sigma_{0}$, the dust surface density becomes low globally, and 
most of the dust mass is retained in the small particles. 
As a result, the whole disk, even the dust trapping region, becomes optically thin.  
At the same time, due to the lower fragmentation barrier $a_{\rm max}$, which linearly 
depends on the gas surface density $\Sigma_{\rm g}$, the maximum dust size also 
decreases with the decreasing $\Sigma_{0}$. In this case, not only the dust continuum flux 
decreases with $\Sigma_{0}$,  but also the global spectral index $\alpha_{\rm 1.0-3.0mm}$ 
increases to values reflecting the dust opacity index $\beta$ of grains with sizes 
much smaller than $\sim 1$ mm.

\begin{figure}[htbp]
\begin{center}
\includegraphics[width=0.45\textwidth]{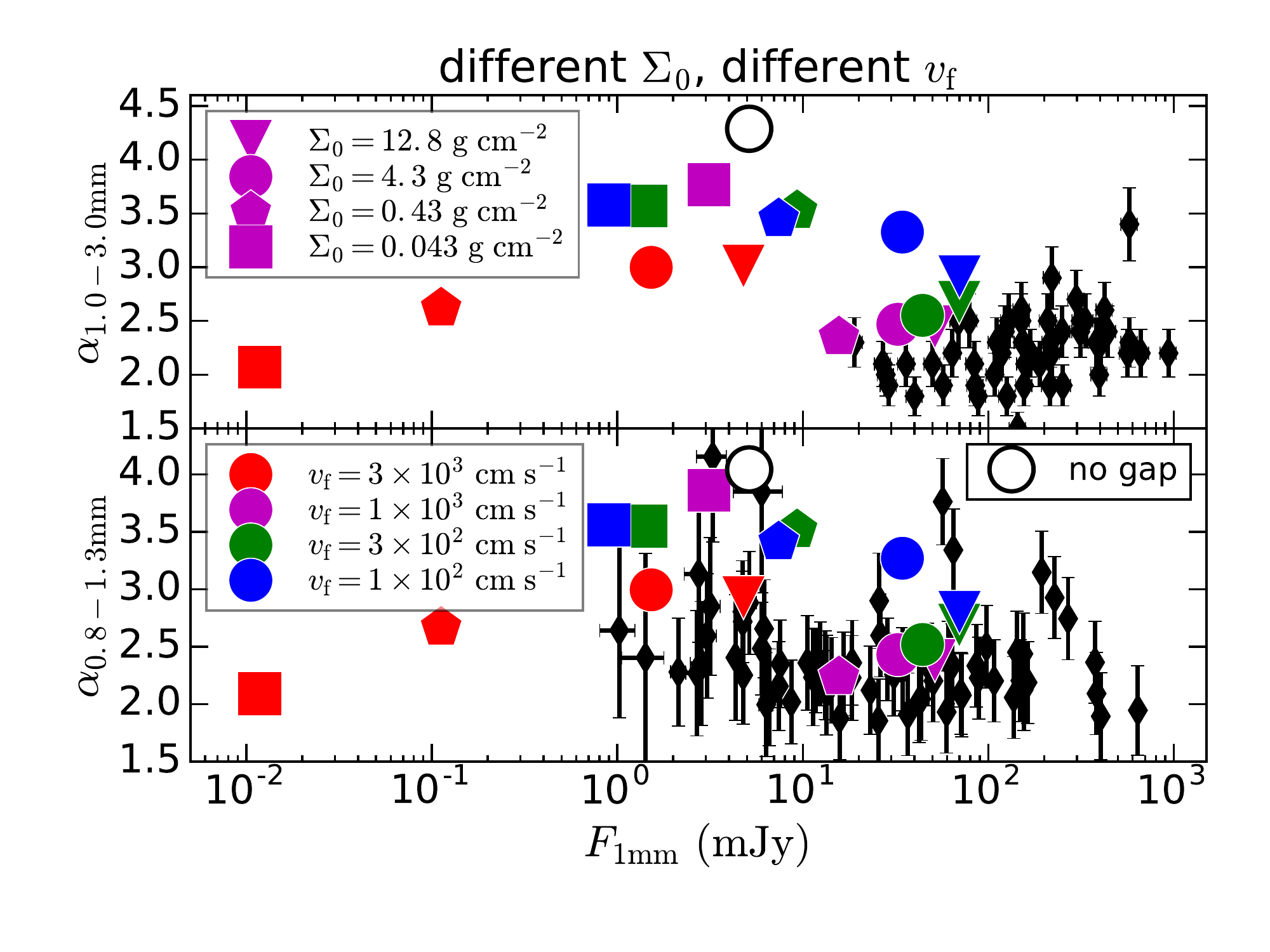}
\end{center}
\caption{Same as Figure~\ref{fig:flux_alpha} but shows the parameter 
dependence of $F_{\rm 1mm}$ and the mm spectral index $\alpha_{\rm mm}$ 
for all the runs with different gas surface densities $\Sigma_{0}$ and 
fragmentation velocities $v_{\rm f}$. The observed data points marked 
as black symbols are the same as Figure~~\ref{fig:fa_m1v2}. 
Symbols with different shapes correspond to different $\Sigma_{0}$, 
while different colors represent different $v_{\rm f}$. 
}\label{fig:vf_sigma}
\end{figure}

We have also made a comprehensive parameter study for the combined effects of different 
$v_{\rm f}$ and $\Sigma_{0}$, and list their results in Table~\ref{tab:para} and
Figure \ref{fig:vf_sigma}.

We find several trends that are described below. For a fixed $v_{\rm f}$, the dependence of $\alpha_{\rm mm}$ on $\Sigma_{0}$ is as follows. When $v_{\rm f}$ is as high as  $3\times10^{3}~{\rm cm\ s^{-1}}$, $\alpha_{\rm mm}$ is insensitive to $\Sigma_{0}$ as the fragmentation barrier $a_{\rm max}$ is already larger than 1 cm even for the lowest values of $\Sigma_{0}$ explored in this study. However, the global spectral index is usually larger than $2.5$, except for extremely low mass disk (i.e., model m3v1). Such a low mass disk is below the detection limit of current observations. While for the case of low $v_{\rm f}$, $\alpha_{\rm mm}$ (1 mm flux $F_{\rm 1mm}$) is anti-correlated (correlated) with $\Sigma_{0}$ due to the optical depth effect. Based on our parameter survey for different $\Sigma_{0}$ and $v_{\rm f}$ listed in Table~\ref{tab:para}, the tentative anti-correlation between $\alpha_{\rm mm}$ and $F_{\rm 1mm}$ found by \citep{Ansdell2018} for some faint sources, as well as the low spectral index for some slightly bright disks from different surveys \citep{Ricci2012,Ansdell2018}, is consistent with a fragmentation velocity in the range of $3\times10^{2}\sim1\times10^{3}~{\rm cm\ s^{-1}}$. This conclusion still holds when we use the opacity from \citet{Semenov2003}. As there are still some uncertainties for the dust opacity and disk parameters (the viscosity profile and the associated gas bump structures as we will discuss later), we should point out that such a constrain on the fragmentation velocity could be not so stringent.

To summarize, both a high fragmentation velocity $v_{\rm f}$ and 
high dust mass can produce a global spectral index $\alpha_{\rm mm}$ 
smaller than 2.5, which are consistent with the mm observations for some 
young PPDs \citep{Ricci2012,Ansdell2018}, as shown in Figure~\ref{fig:flux_alpha}. 
Such a model degeneracy can be broken by checking the variation of 
the spectral index radial profile with different wavelengths, which 
will be discussed in Section~\ref{sec:rad_profile}. 
When both the  fragmentation velocity and dust mass are below a critical value 
such that the fragmentation barrier $a_{\rm max}<1\ {\rm mm}$  and the 
effective optical depth $\tau_{\rm eff}\ll1$, the global spectral index $\alpha_{\rm mm}>3.0$.

\subsubsection{Viscosity}

The viscosity is expected to influence the gas bump properties and modify the dust dynamics as well.
We simulate several models with different viscosity parameters $\alpha_{0}$. 
The spectral index $\alpha_{\rm mm}$ at different bands and 1 mm flux are listed in Table~\ref{tab:para}
and are shown in Figure~\ref{fig:flux_alpha}.  

\begin{figure}[htbp]
\begin{center}
\includegraphics[width=0.45\textwidth]{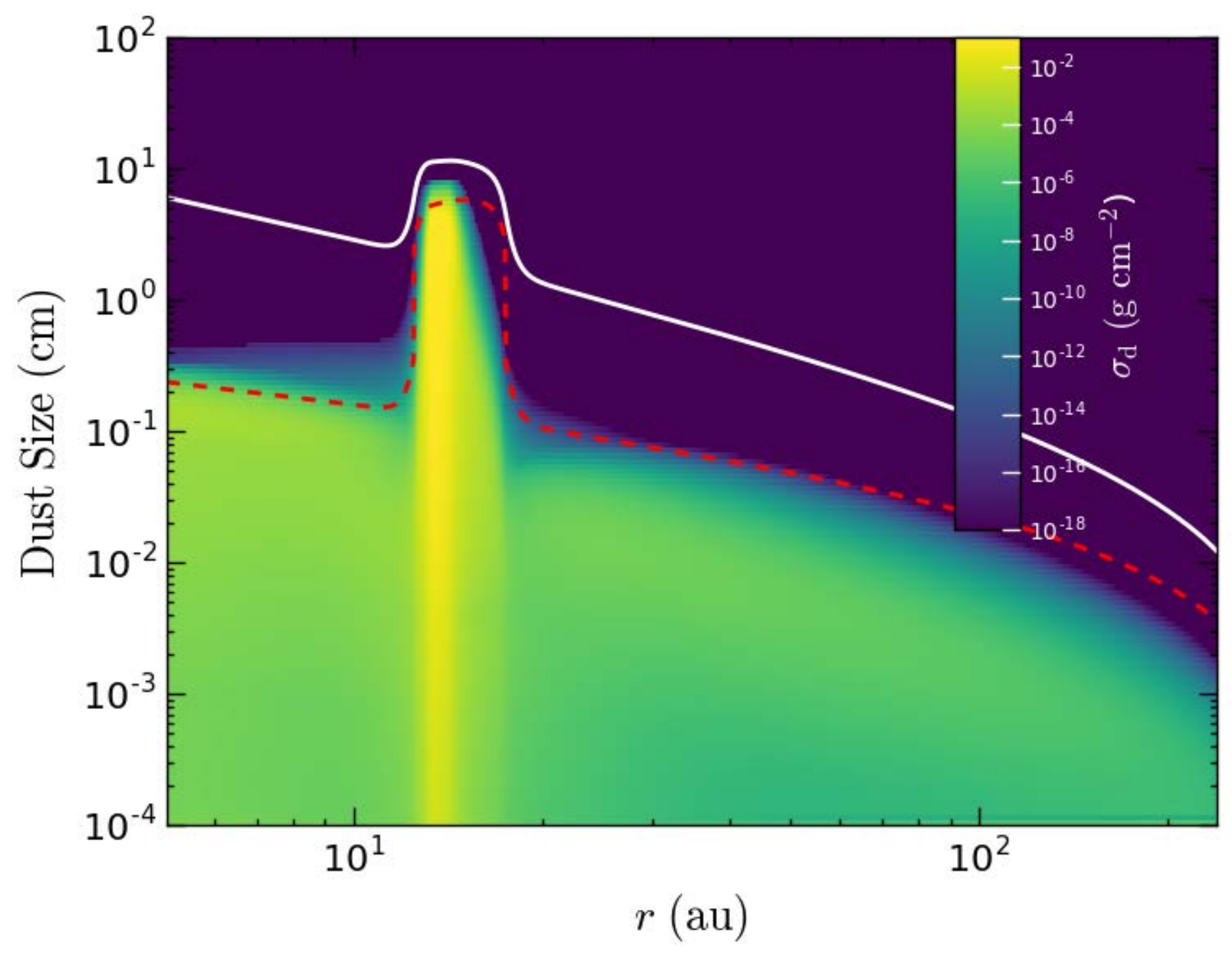}
\end{center}
\caption{Same as Figure~\ref{fig:surf_dust} but for model m1a1 where the disk viscosity is 
a factor of 2 lower.}\label{fig:surf_dust_m1a1}
\end{figure}

According to Equation
(\ref{eq:amax_frag}), the maximum dust size increases as viscosity decreases. This affects the optical depth
significantly. Indeed, as shown in Figure~\ref{fig:surf_dust_m1a1} where the run is similar to our
fiducial run except that $\alpha_{0} =5\times10^{-3}$ is a factor of 2 lower, the fragmentation barrier 
in the whole disk $a_{\rm max}$ is larger. This leads to two consequences. One is that 
$F_{\rm 1mm}$ decreases with decreasing $\alpha_{0}$ because the dust in the whole disk can 
coagulate to a larger size thanks to the larger fragmentation barrier, which results in the emission 
shifting from the optically thick to thin regime for most of the disk region. The other is that 
$\alpha_{\rm mm}$ becomes increasingly larger. It turns out that the dust bump region is still optically
thick and it gives a spectral index around $2.0$, but the flux aside from the bump region 
can contribute a considerable fraction of the total emission. This flux has 
a spectral index close to $2.5\sim3.5$ because of the optically thin emission. 
The global spectral index in the whole disk then gives a value of about $3.0$.

In order to produce brighter disks, a relatively high viscosity parameter $\alpha_{0}$ is thus required.
Note that the increase of global spectral index $\alpha_{\rm mm}$ is mainly attributed to the 
global increase of the fragmentation barrier $a_{\rm max}$.
In order to produce a relatively low spectral index with a low disk viscosity, 
we need to adopt a lower fragmentation velocity $v_{\rm f}$ to limit the dust sizes.

\subsubsection{Viscosity Gap Depth, Width and Location}

We now study how viscosity gap depth affects the emission properties.
We used three $d_{\rm ring} = 2.0, 10.0, 20.0$ with fixed $\alpha_{0}$, which are
shown as the models m1d1, m1d2, and m1d3 in Table~\ref{tab:para} and Figure~\ref{fig:flux_alpha}.  

Overall, $\alpha_{\rm mm}$ is insensitive to $d_{\rm ring}$.
This is because the emission in the trapping region can either be optically thick or 
dominated by larger dust particles, both of which lead to a shallow spectral shape, 
while the emission outside the bump region does not vary significantly.
Furthermore, we find that even a slightly shallower gap for the viscosity profile, i.e., $d_{\rm ring}=2.0$, 
can still trap enough dust and result in dust size growing. The value of global spectral index 
$\alpha_{\rm 0.8-1.3mm}$ increases slightly, but is still smaller than 3.0. 
In order to reproduce a spectral index lower than 2.5, the gap depth for the gas viscosity, 
or equivalently the gas density contrast, should be larger than $\sim 3$.

In the situation where the gas bump is produced via a mechanism different from a viscosity gap, such a gap should not be used in the coagulation/fragmentation calculation. This is because the real viscosity should be smooth in the whole disk, and our viscosity gap is only used for generating the bump. This has been tested for our fiducial model. We find that the maximum dust size reached in the bump region decreases by a factor of 6, which is $\sim2\ {\rm mm}$. The mm flux and the spectral index are 42.2 mJy and 2.46, respectively, which are actually close to our fiducial model results.

We have also adjusted the gap width by using different $w_{\rm ring}$, 
$2.5$ au, $7.5$ au, and $10.0$ au, of the viscosity profile. 
Results are listed in Table~\ref{tab:para} and shown in Figure~\ref{fig:flux_alpha}.
We find that the final spectral slopes and $F_{\rm 1mm}$
are insensitive to changes of $w_{\rm ring}$ 
between $2.5~{\rm au}$ and $10~{\rm au}$. 

The dependence of disk emissions on the ring location is investigated by placing the 
viscosity gap at $r_{\rm ring} =$ $25~{\rm au}$, $50~{\rm au}$, and $75$ au. The results are shown in Figure~\ref{fig:flux_alpha}.
In general, the 1 mm flux increases slightly and the spectral slope increases 
when $r_{\rm ring}$ moves outward. This is because the dust mass interior to the
dust bump region, $r\lesssim r_{\rm ring}-w_{\rm ring}/2$, gets larger. 
This region has a relatively steep spectral slope because it is optically thin emission from small dust grains. 
The emission from the dust trapping region can still be optically thick. 
Therefore, as $r_{\rm ring}$ increases, both the global spectral index and the total disk flux increase. 
As shown in Figure~\ref{fig:flux_alpha}, the global spectral indices from these models give 
somewhat higher values than the observations.

\subsubsection{Gas Scale Height and Surface Density Profile}

We further consider two other disk parameters, gas scale height $h_{0}$ and 
gas surface density profile $\gamma$. The gas scale height can influence the gas 
temperature (pressure) profile. We show one example for the variation of the mid-plane 
dust temperature with the disk scale height in Figure~\ref{fig:temp}. 
The temperature is slightly higher because of a higher sound speed.  
The $\gamma$ parameter can change the total amount of dust. 
The models labeled as m1h1(2) and m1b1(2) with the corresponding results are shown in 
Table~\ref{tab:para}. For all these models, the spectral index and mm flux variations are only modest. 
This is because the dust trapping region is mostly optically thick, and in the optically thick regime 
the emission depends on temperature, not on the grain size distribution. 
The small variation of the flux and spectral index is due to the small variation in 
the dust temperature in the dust trapping region as shown in Figure~\ref{fig:temp}.

\subsubsection{No Viscosity Gap -- No Ring}

\begin{figure*}[htbp]
\begin{center}
\includegraphics[width=0.45\textwidth]{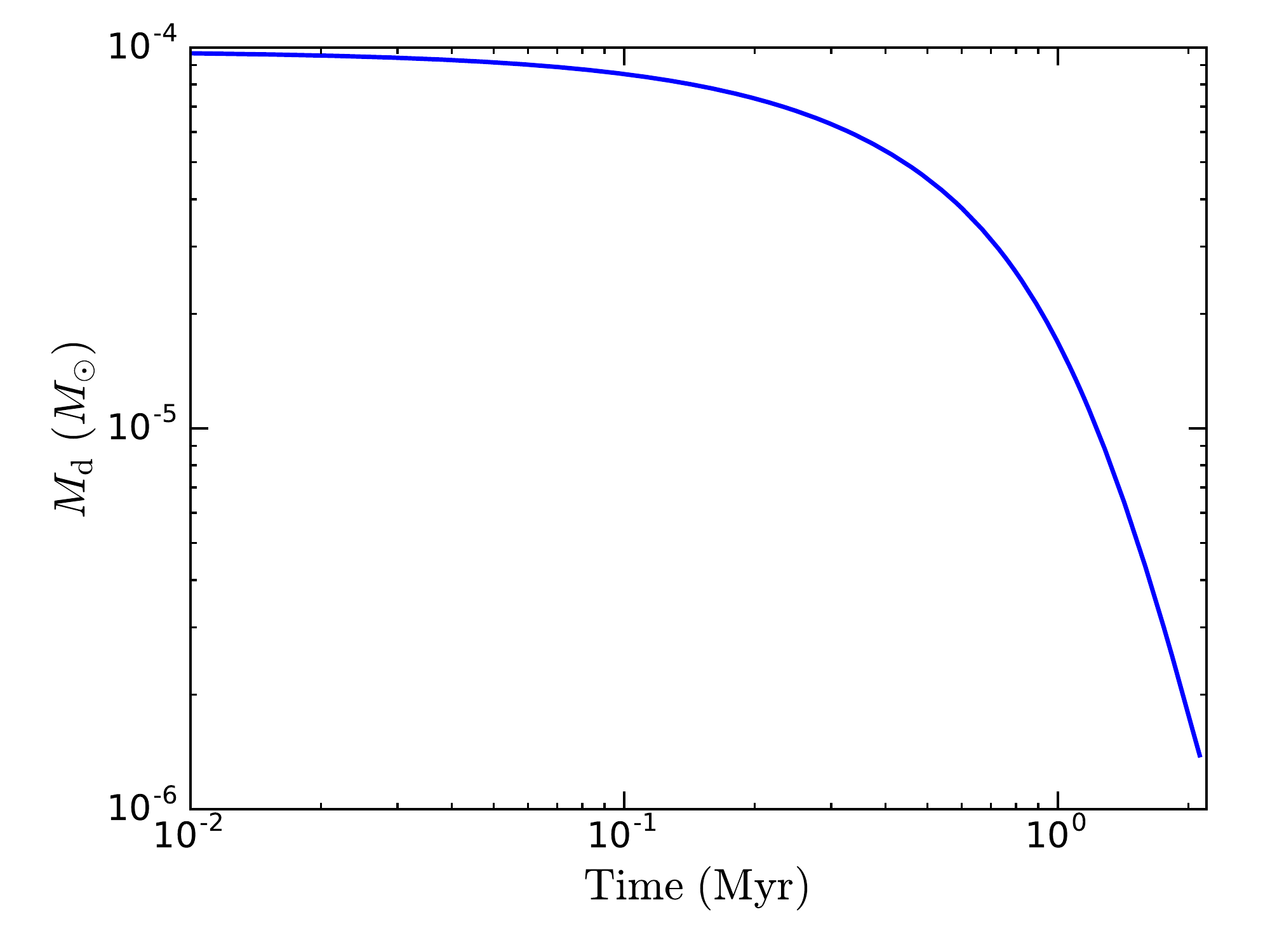}
\hskip 0.5truecm
\includegraphics[width=0.45\textwidth]{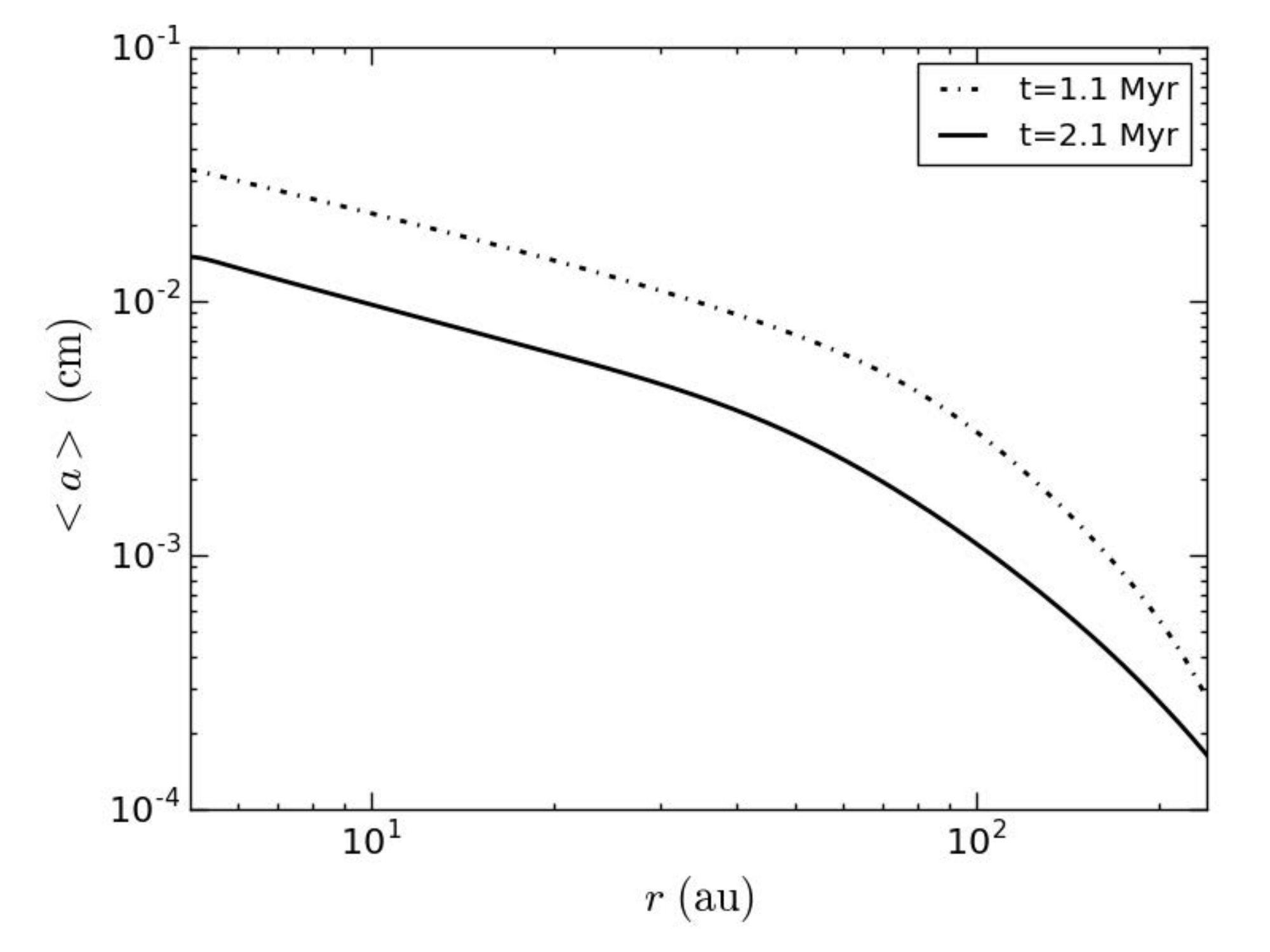}
\end{center}
\caption{Model without any viscosity gap (model m1ng). Left panel: the time evolution of 
the total dust mass in the whole disk. Right panel: grain size 
averaged by the dust surface density at each radius, $\langle a \rangle$, at two different epochs. 
Other disk parameters are listed in Table~\ref{tab:para}.}\label{fig:m1ng}
\end{figure*}

In order to quantify the role of the viscosity gap/ring, we study the case with 
$\alpha_{\rm vis}=\alpha_{0}=10^{-2}$ throughout the whole disk, and all the other disk parameters 
the same as in our fiducial model m1v2. At the end of the 2.1 Myr evolution, the total dust mass is 
only  $1.3\times10^{-6}M_{\odot}$, or $\le1\%$ of the original dust is retained in the disk and is
two orders of magnitude smaller than for model m1v2, 
as shown in the left panel of Figure~\ref{fig:m1ng}. 
In addition, low dust surface density severely limits the ability of dust particles growing 
to larger sizes, as shown in the right panel of Figure~\ref{fig:m1ng}. 
In fact, almost all the bigger grains will drift radially inward due to their larger Stokes number. 

We then repeat the same radiative transfer calculations to produce the dust continuum 
emission for this disk model.
The obtained values for the spectral index for the emission within 100 au 
from the star are quite large, reaching values of $\approx4.0$ 
due to an inefficient size growth and an optically thin emission for the dust. 
As expected, the flux at 1 mm is only 5.0 mJy due to the low total dust mass. 
It is difficult to decrease the global spectral index down to the observed values by changing 
fragmentation velocity and/or initial dust mass because the disk emission is dominated by 
sub-mm sized particles in the outer region of the disk.

\begin{figure*}[htbp]
\begin{center}
\includegraphics[width=0.45\textwidth]{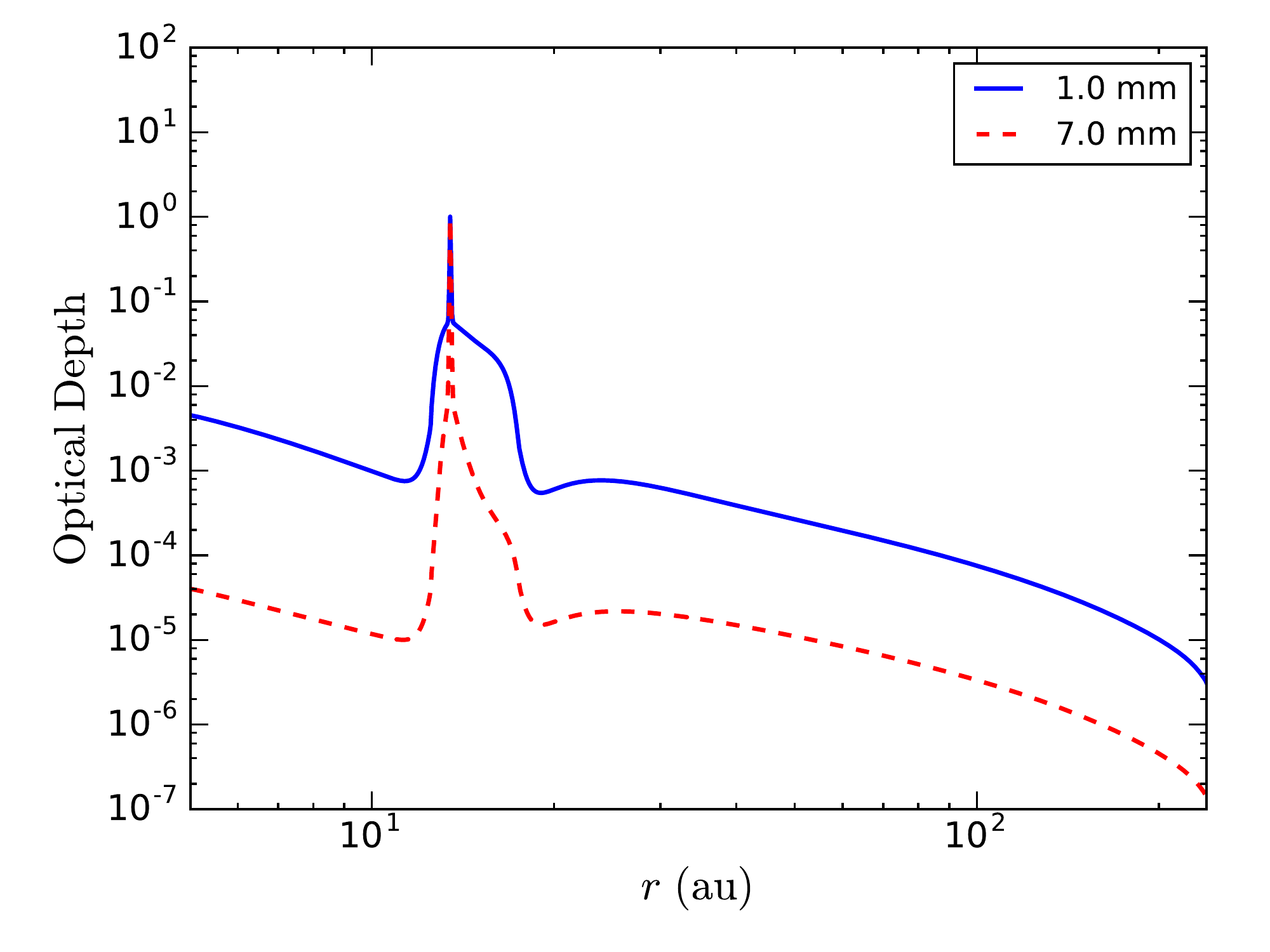}
\includegraphics[width=0.45\textwidth]{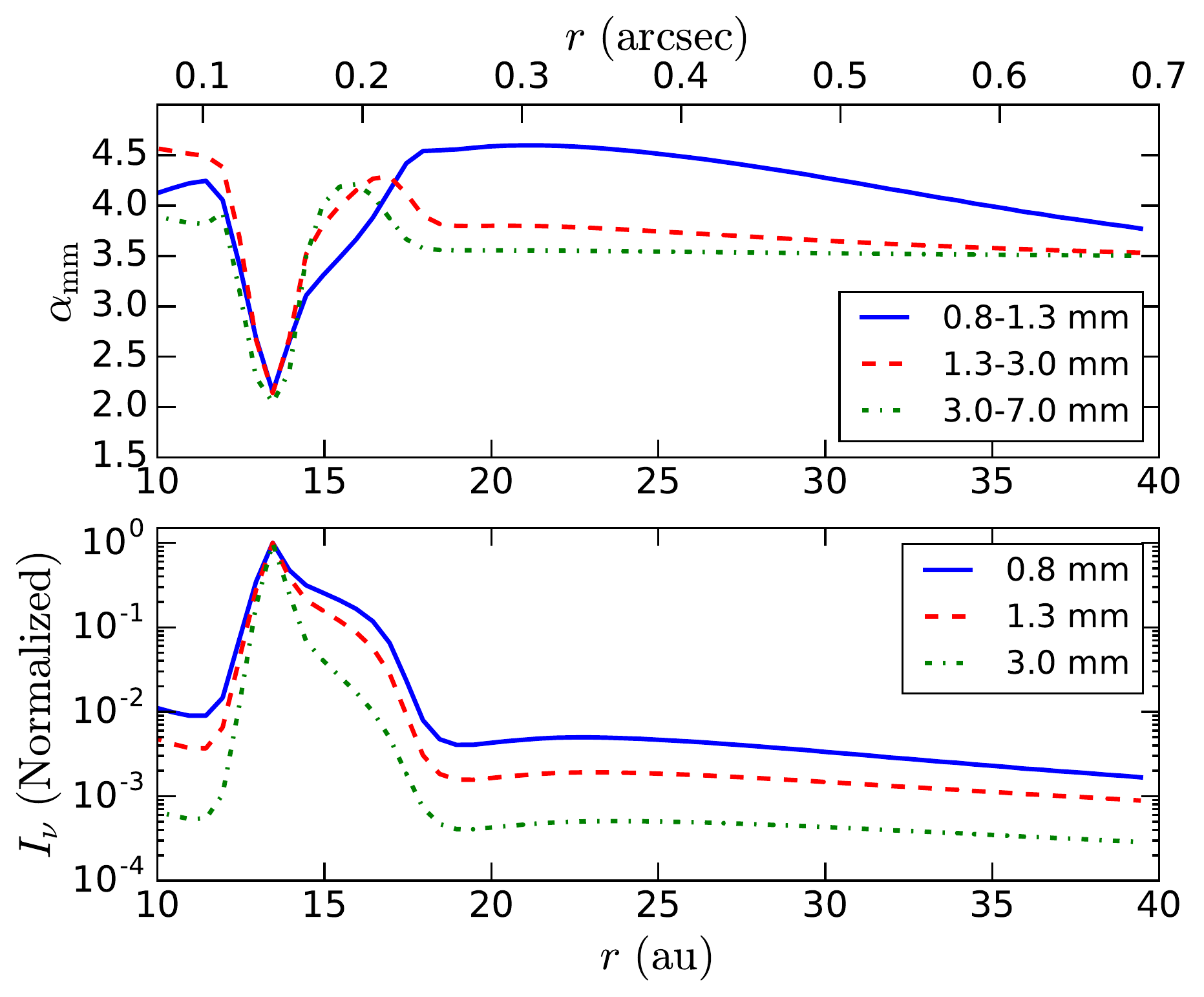}
\includegraphics[width=0.45\textwidth]{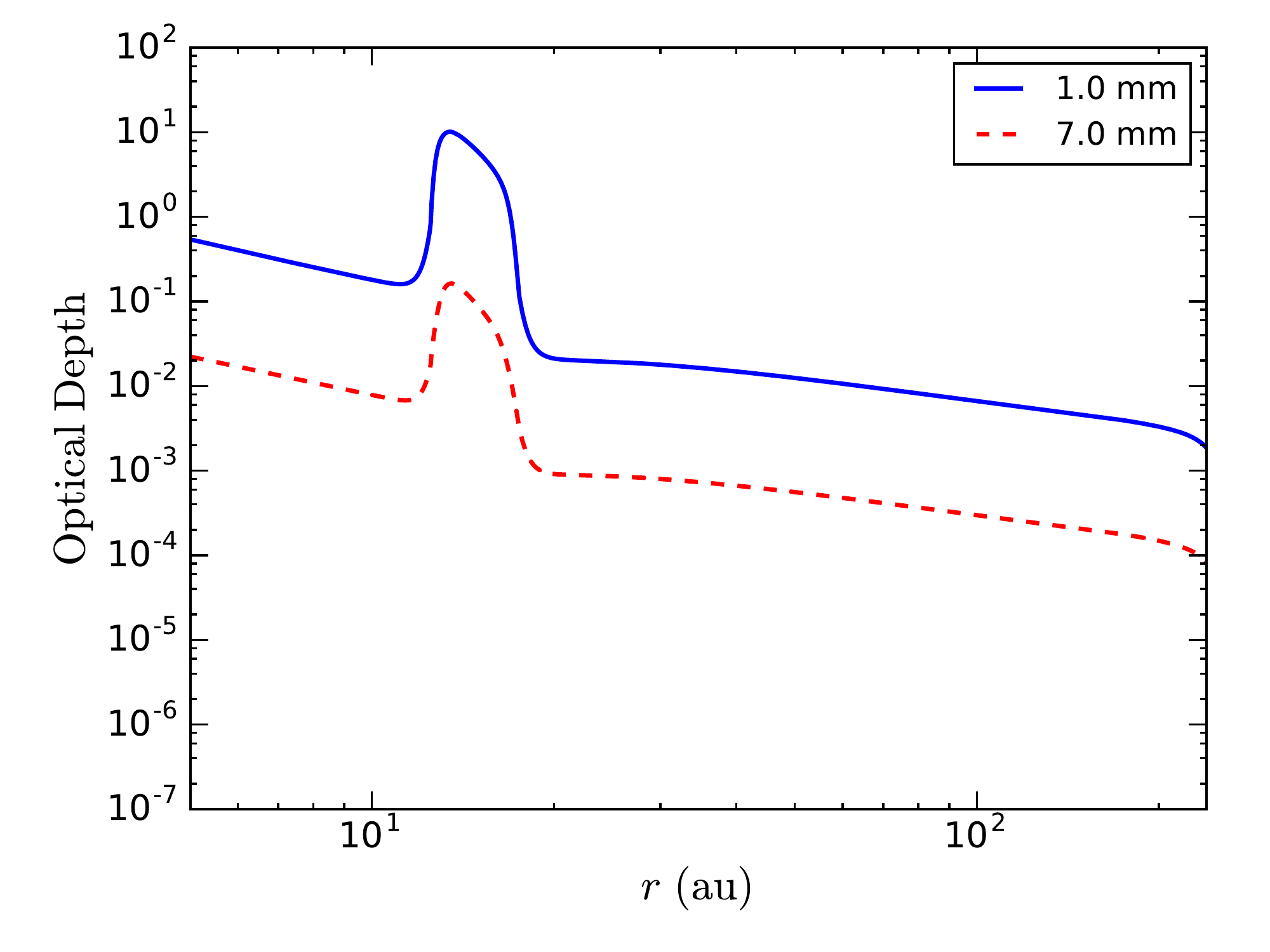}
\includegraphics[width=0.45\textwidth]{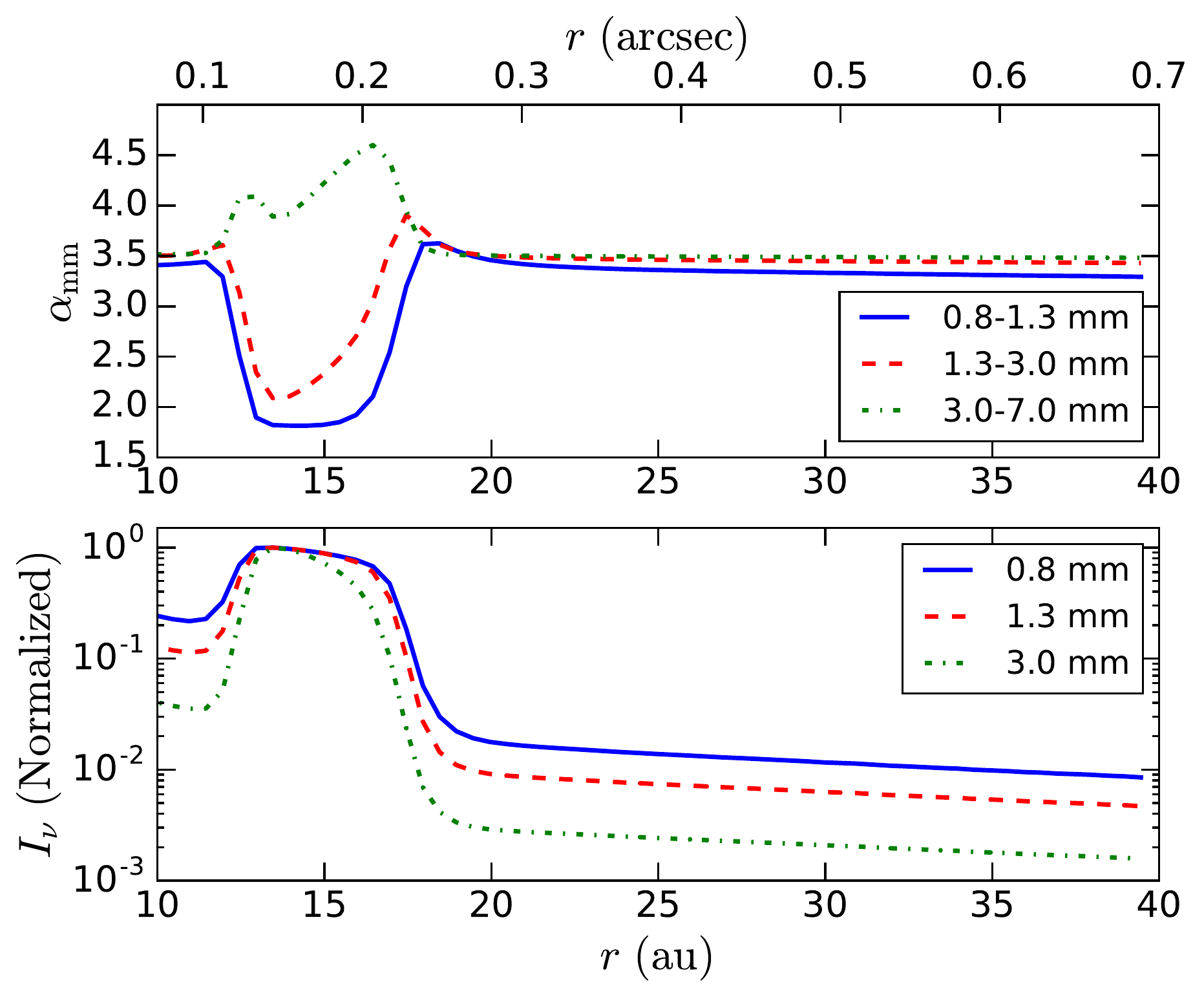}
\end{center}
\caption{Similar to Figure~\ref{fig:optical_depth}, but for model m1v1 (upper panel) and 
model m1v3 (lower panels). Here we only show the optical depth profile close to the end of the simulation, i.e., 2.1 Myr.}\label{fig:rad_spec}
\end{figure*}

\subsection{Differentiating Effects of Optical Depth and Dust Size}
\label{sec:rad_profile}

Both optically thick emission and efficient dust growth can be responsible for the 
low values of the global spectral indices, and one key question is how to discriminate 
between these two scenarios. We resort to the spatially resolved radial profile of the 
disk emission at multiple wavelengths to break such degeneracy.

To quantify our method, we choose two models with different values of fragmentation velocity $v_{\rm f}$ to compare with our fiducial model m1v2. We convolve the image produced from \texttt{RADMC-3D} with a Gaussian beam of the FWHM 
$0.003^{\prime\prime}\times0.003^{\prime\prime}$, close to the full resolution of 
our \texttt{RADMC-3D} grid, which can recover all the substructures produced in the 
hydrodynamical model. Although such resolution is beyond the current limit of ALMA observations, 
our goal is to diagnose the physical processes that contribute to 
different spectral shapes of the dust emission.

The radial profile of flux and spectral index at mm wavelengths for model m1v1 
(high $v_{\rm f}$) are shown in the upper panels of Figure~\ref{fig:rad_spec}.
Most of the disk is optically thin for both 1 mm and 7 mm, which is responsible for
the relatively high spectral index throughout the disk. There exists a
narrow region within the dust ring where it is marginally optically thick and 
the spectral index dips to $\sim2.0$ at that location. 
This is caused by the biggest particles which dominate the emission 
have grown to be much larger than $\sim7/2\pi\ {\rm mm}\simeq1.0\ {\rm mm}$.
The sharp peak near the inner edge of the dust ring in the optical depth profile is contributed to the large particles which are sharply trapped there.
The overall radial profile of spectral index is somewhat similar to model m1v2 case except that the gap in the viscosity profile becomes narrower, especially at the longer wavelength, as shown in the upper panel of Figure~\ref{fig:rad_spec}.

For a lower fragmentation velocity model m1v3, the emission can 
be optically thick at 1 mm, but optically thin at 7 mm band in the dust trapping region, 
as shown in the lower left panel of Figure~\ref{fig:rad_spec}. This is also contrary 
to the case of m1v2 shown in the upper panel of Figure~\ref{fig:optical_depth}, 
where the emission in the dust trapping region is optically thick both at 1 mm and 7 mm. 
We show the corresponding radial profile of the spectral index and flux 
in the lower right panel of Figure~\ref{fig:rad_spec}. The lowest spectral index 
between $0.8-1.3$ mm is 
about 2.0, but at longer wavelengths, e.g. at $3.0-7.0$ mm, 
it reaches much higher values in the same region. This reason is as follows. 
First,  this is due to the optically thin emission at longer wavelength 
as shown in the lower left panel of Figure~\ref{fig:rad_spec}. Second, 
in the optically thin regime, the grains cannot grow to a size larger than 
$1.0\ {\rm mm}$ because of the lower fragmentation velocity. 

As a comparison to our fiducial model with $v_{\rm f}=1\times10^{3}\ {\rm cm\ s^{-1}}$ 
(model m1v2), the spectral index at the dust trapping region can reach 2.0 both 
at short waveband ($0.8-1.3$ mm) and long waveband ($3.0-7.0$ mm), 
which is a consequence of the large optical depth.

As shown in these cases, the different radial behavior of the spectral index 
at multiple mm wavelengths can be used to break the model degeneracy 
as well as set some constraints on the dust size\footnote{These constraints 
will be affected by the uncertainties in the dust opacity or the unknown porosity of the particles.}. 
This can be tested by future high resolution observations of ALMA and ngVLA at longer wavelengths.

\section{Summary and Discussions} \label{sec:conclusions}

In this work, we present detailed 1D hydrodynamical simulations of PPDs 
to study the two-fluid (gas+dust) co-evolution with a state-of-the-art dust evolution code 
\citep{Li2005,Li2009,Fu2014,Birnstiel2010a} with dust size growth via coagulation and fragmentation 
\citep{Birnstiel2010a}. This allows us to self-consistently 
obtain the quasi-steady state gas and dust distributions including dust feedback, 
dust radial drift, and coagulation/fragmentation balance.  
The comparisons with observations are then made using the \texttt{RADMC-3D} 
code for dust continuum radiative transfer. 

By assuming a viscosity gap in the inner region of the disk ($\sim 15$ au), 
a gas bump as well as a dust trap are produced. With a systematic parameter study for 
different disk gas and dust parameters and coagulation velocities, we find:
\begin{itemize}
\item A gas and dust ring at the inner disk can produce a lower global spectral index. 

\item High fragmentation velocities $v_{\rm f}$ ($1000~ {\rm cm~s^{-1}}\lesssim 
v_{\rm f}\lesssim3\times 10^3~{\rm cm~s^{-1}}$) can facilitate efficient 
dust size growth (Figure~\ref{fig:surf_dust}), which produces a dust mm 
spectral index $\alpha_{\rm mm}$ close to 2.0, even for optically thin emission. 
For lower values of $v_{\rm f}$ (300 to $1000~{\rm cm~s^{-1}}$), 
a high dust surface density in the dust trapping region leads to optically thick emission, 
which then results in a low $\alpha_{\rm mm}$ as well (Figure~\ref{fig:fa_m1v2}, Figure~\ref{fig:flux_alpha}).

\item The high flux and low spectral index sources from a few surveys \citep{Ricci2012,Ansdell2018} can be explained by a fragmentation velocity $v_{\rm f}$ in the range $300 - 1000~{\rm cm~s^{-1}}$. The tentatively negative relation between $\alpha_{\rm mm}$ and $F_{\rm 1mm}$ shown from \citet{Ansdell2018} for the faint PPD sources can also be interpreted by a relatively low $v_{\rm f}$ ($\lesssim1000~{\rm cm~s^{-1}}$) with a variation of dust mass surface density. This is simply because the optical depth becomes smaller as the dust mass, and mm flux, decrease (Figure~\ref{fig:flux_alpha}).

\item The spectral index at mm wavelengths $\alpha_{\rm mm}$ is not very sensitive to the gas bump width $w_{\rm ring}$, gap depth $d_{\rm ring}$,  disk scale height $h_{0}$, and disk profile index $\gamma$. A very low global disk viscosity $\alpha_{0}\lesssim10^{-3}$ cannot reproduce the observed low global spectral index. 

\item The one-ringed disk models can reproduce the observed properties of the faint part of the observed disks ($\lesssim 100$ mJy) reasonably well, but they cannot reproduce the brightest disks in the observational surveys (Figure~\ref{fig:flux_alpha}). We expect that multiple rings could be responsible for these bright disks as discussed later.

\item While both large size grains and optically thick emission contribute to a small 
spectral index $\alpha_{\rm mm}\lesssim2.5$, the future high resolution observations, 
e.g., by ALMA and ngVLA, can distinguish these two effects based on their spatial distribution properties 
(Figure~\ref{fig:optical_depth}, Figure~\ref{fig:rad_spec}).

\end{itemize}

In this work, we adopt a viscosity parameter $\alpha_{\rm 0}=10^{-2}$ for most of our disks. This value is a factor of a few higher than some observational constrained values based on the vertical thinness of some disks with rings and gaps, e.g., HL Tau (\citealt{Pinte2016}; $\alpha_{\rm vis}\sim3\times10^{-4}$) and HD 163296 (\citealt{Flaherty2017}; $\alpha_{\rm vis}\lesssim3\times10^{-3}$).  On the one hand, such a global low $\alpha_{\rm vis}$ could reflect the low turbulence in the ringed/gap regions. Therefore, it indicates that some regions in the disk could be indeed magnetorotational instability (MRI) inactive compared with the theoretical expectation of MRI. On the other hand, for a lower viscosity $\alpha_{\rm vis}$ in a very thin disk, we find that an even lower fragmentation velocity ($\lesssim3\ {\rm m\ s^{-1}}$) is required to produce a low spectral index ($\alpha_{\rm mm}\lesssim2.5$). One issue we should point out is that we keep the viscosity profile $\alpha_{\rm vis}(r)$ independent on dust evolution. This is contrary to the expectation of the MRI dead zone scenario. \citet{Dzyurkevich2013} has shown that $\alpha_{\rm vis}$ depends on ionization fraction, which depends on dust properties. 

We have ignored the dust scattering effect in the radiative transfer. Recently, \citet{Zhu2019} (see also \citealt{Liu2019}) found that dust scattering can reduce the emission from an optically thick region. It leads to the apparent interpretation that bright disks tend to be harder. It can also modify the spectral slope at mm band depending the variation of the dust albedo with wavelength. However, the modification of the spectral shape is not significant (close to the expectation of 2.0) if the dust size is much larger than $\lambda/2\pi$, which is the case in the dust trapping region in most of our models. This issue will be addressed with detailed radiative  transfer calculations in the future.

We have also explored the effects of multiple rings. 
A very low global spectral index (smaller than 2.5) can be obtained with a reasonable 
combination of disk parameters and fragmentation velocity ($\sim10\ {\rm m\ s^{-1}}$), 
and more importantly, the disk becomes systematically brighter. 
Multiple rings can increase the fraction of optically thick emission regions, 
thus significantly increase the disk total flux, while still keeping low values for the global 
spectral indices around mm wavelength. These are consistent with recent ALMA 
surveys for 12 disks in Taurus star-forming region \citep{Long2018} and 18 single-disk systems 
in the DSHARP program \citep{Andrews2018DSHARP,Huang2018DSHARP}, where disks with multiple sub-structures tend to be brighter, although an observational bias might exist.

The ringed structures we inferred could be the location to trigger further planetesimal formation. 
If the ringed structure is associated with the MRI dead  zone, the less turbulence in the gas bump would result in lower collision speeds for particles, which would be more suitable for further grain growth. Such a gas bump can also slow down the rapid radial drift to allow more time for planetesimal formation. In addition, the concentration of particles in high densities can trigger planetesimal formation via streaming instability \citep{Youdin2005} and the gravitational collapse \citep[e.g., see review by][]{Pinilla2017}. The planetesimal formation process can remove some mm-size dust particles, and hence reduce the mm flux of the disk, which makes our model even harder to explain the brighter disks.

\acknowledgments

We would like to thank Ruobing Dong and Pinghui Huang for helpful discussions, and the referee for useful comments on the paper. YPL, HL and SL gratefully acknowledge the support by LANL/CSES and NASA/ATP.  
T.B. acknowledges funding from the European Research Council (ERC) under the European Union's Horizon 
2020 research and innovation programme under grant agreement No 714769 and funding by the Deutsche 
Forschungsgemeinschaft (DFG, German Research Foundation) Ref no. FOR 2634/1. FY acknowledges the support by Natural Science Foundation of China under grant 11661161012. Simulations of this work 
were performed with LANL Institutional Computing.

\software{\texttt{Astropy} \citep{Astropy2013},
          \texttt{LA-COMPASS} \citep{Li2005,Li2009},
          \texttt{Matplotlib} \citep{Hunter2007},
          \texttt{Numpy} \citep{vanderWalt2011},
          \texttt{RADMC-3D} \citep{Dullemond2012}
          }


\begin{thebibliography}{}

\bibitem[ALMA Partnership et al.(2015)]{ALMA2015} ALMA Partnership, Brogan, C.~L., P{\'e}rez, L.~M., et al.\ 2015, \apjl, 808, L3

\bibitem[Andrews(2015)]{Andrews2015} Andrews, S.~M.\ 2015, \pasp, 127, 961

\bibitem[Andrews et al.(2018)]{Andrews2018DSHARP} Andrews, S.~M., Huang, J., P{\'e}rez, L.~M., et al.\ 2018, \apj, 869, L41.

\bibitem[Andrews et al.(2010)]{Andrews2010} Andrews, S.~M., Wilner, D.~J., Hughes, A.~M., Qi, C., \& Dullemond, C.~P.\ 2010, \apj, 723, 1241

\bibitem[Andrews et al.(2016)]{Andrews2016} Andrews, S.~M., Wilner, D.~J., Zhu, Z., et al.\ 2016, \apjl, 820, L40

\bibitem[Ansdell et al.(2018)]{Ansdell2018} Ansdell, M., Williams, J.~P., Trapman, L., et al.\ 2018, \apj, 859, 21

\bibitem[Armitage(2010)]{Armitage2010} Armitage, P.~J.\ 2010, Astrophysics of Planet Formation, by Philip J.~Armitage, 294 pp.~ISBN 978-0-521-88745-8 (hardback).~Cambridge, UK: Cambridge University Press, 2010.,

\bibitem[Armitage(2011)]{Armitage2011} Armitage, P.~J.\ 2011, \araa, 49, 195

\bibitem[Astropy Collaboration et al.(2013)]{Astropy2013} Astropy Collaboration, Robitaille, T.~P., Tollerud, E.~J., et al.\ 2013, \aap, 558, A33

\bibitem[Balbus \& Hawley(1991)]{Balbus1991} Balbus, S.~A., \& Hawley, J.~F.\ 1991, \apj, 376, 214

\bibitem[Birnstiel et al.(2010a)]{Birnstiel2010a} Birnstiel, T., Dullemond, C.~P., \& Brauer, F.\ 2010a, \aap, 513, A79

\bibitem[Birnstiel et al.(2018)]{Birnstiel2018} Birnstiel, T., Dullemond, C.~P., Zhu, Z., et al.\ 2018, \apjl, 869, L45

\bibitem[Birnstiel et al.(2012)]{Birnstiel2012} Birnstiel, T., Klahr, H., \& Ercolano, B.\ 2012, \aap, 539, A148

\bibitem[Birnstiel et al.(2010b)]{Birnstiel2010b} Birnstiel, T., Ricci, L., Trotta, F., et al.\ 2010b, \aap, 516, L14

\bibitem[Bjorkman \& Wood(2001)]{Bjorkman2001} Bjorkman, J.~E., \& Wood, K.\ 2001, \apj, 554, 615

\bibitem[Blum \& Wurm(2008)]{Blum2008} Blum, J., \& Wurm, G.\ 2008, \araa, 46, 21

\bibitem[Boehler et al.(2018)]{Boehler2018} Boehler, Y., Ricci, L., Weaver, E., et al.\ 2018, \apj, 853, 162

\bibitem[Boss(2010)]{Boss2010} Boss, A.~P.\ 2010, \apjl, 725, L145

\bibitem[Brauer et al.(2008)]{Brauer2008} Brauer, F., Dullemond, C.~P., \& Henning, T.\ 2008, \aap, 480, 859

\bibitem[Brauer et al.(2007)]{Brauer2007} Brauer, F., Dullemond, C. P., Johansen, A., Henning, Th., Klahr, H., \& Natta, A.\ 2007, \aap, 469, 1169

\bibitem[Bryden et al.(1999)]{Bryden1999} Bryden, G., Chen, X., Lin, D.~N.~C., Nelson, R.~P., \& Papaloizou, J.~C.~B.\ 1999, \apj, 514, 344

\bibitem[Burke et al.(2015)]{Burke2015} Burke, C. J., Christiansen, J. L., Mullally, F., Seader, S., Huber, D. et al.\ 2015, \apj, 809, 8

\bibitem[Chiang \& Murray-Clay(2007)]{Chiang2007} Chiang, E., \& Murray-Clay, R.\ 2007, Nature Physics, 3, 604

\bibitem[Chiang \& Youdin(2010)]{Chiang2010} Chiang, E., \& Youdin, A.~N.\ 2010, Annual Review of Earth and Planetary Sciences, 38, 493


\bibitem[Cieza et al.(2017)]{Cieza2017} Cieza, L.~A., Casassus, S., P{\'e}rez, S., et al.\ 2017, \apjl, 851, L23

\bibitem[Cox et al.(2017)]{Cox2017} Cox, E.~G., Harris, R.~J., Looney, L.~W., et al.\ 2017, \apj, 851, 83

\bibitem[Dipierro et al.(2015)]{Dipierro2015} Dipierro, G., Price, D., Laibe, G., et al.\ 2015, \mnras, 453, L73

\bibitem[Dipierro et al.(2018)]{Dipierro2018} Dipierro, G., Ricci, L., P{\'e}rez, L., et al.\ 2018, \mnras, 475, 5296

\bibitem[Dominik \& Tielens(1997)]{Dominik1997} Dominik, C., \& Tielens, A.~G.~G.~M.\ 1997, \apj, 480, 647

\bibitem[Dong et al.(2017)]{Dong2017} Dong, R., Li, S., Chiang, E., \& Li, H.\ 2017, \apj, 843, 127

\bibitem[Dullemond et al.(2018)]{Dullemond2018} Dullemond, C.~P., Birnstiel, T., Huang, J., et al.\ 2018, \apjl, 869, L46

\bibitem[Dullemond et al.(2012)]{Dullemond2012} Dullemond, C.~P., Juhasz, A., Pohl, A., et al.\ 2012, Astrophysics Source Code Library, ascl:1202.015

\bibitem[Draine(2006)]{Draine2006} Draine, B.~T.\ 2006, \apj, 636, 1114

\bibitem[Dzyurkevich et al.(2013)]{Dzyurkevich2013} Dzyurkevich, N., Turner, N.~J., Henning, T., \& Kley, W.\ 2013, \apj, 765, 114 

\bibitem[Fedele et al.(2018)]{Fedele2018} Fedele, D., Tazzari, M., Booth, R., et al.\ 2018, \aap, 610, A24

\bibitem[Flaherty et al.(2017)]{Flaherty2017} Flaherty, K.~M., Hughes, A.~M., Rose, S.~C., et al.\ 2017, \apj, 843, 150 

\bibitem[Fromang \& Nelson(2005)]{Fromang2005} Fromang, S., \& Nelson, R.~P.\ 2005, \mnras, 364, L81

\bibitem[Fu et al.(2014)]{Fu2014} Fu, W., Li, H., Lubow, S., Li, S., \& Liang, E.\ 2014, \apjl, 795, L39

\bibitem[Gammie(1996)]{Gammie1996} Gammie, C.~F.\ 1996, \apj, 457, 355

\bibitem[Huang et al.(2018a)]{Huang2018} Huang, J., Andrews, S.~M., Cleeves, L.~I., et al.\ 2018a, \apj, 852, 122

\bibitem[Huang et al.(2018b)]{Huang2018DSHARP} Huang, J., Andrews, S.~M., Dullemond, C.~P., et al.\ 2018b, \apj, 869, L42.

\bibitem[Hunter(2007)]{Hunter2007} Hunter, J.~D.\ 2007, Computing in Science and Engineering, 9, 90

\bibitem[Igea \& Glassgold(1999)]{Igea1999} Igea, J., \& Glassgold, A.~E.\ 1999, \apj, 518, 848

\bibitem[Isella et al.(2009)]{Isella2009} Isella, A., Carpenter, J.~M., \& Sargent, A.~I.\ 2009, \apj, 701, 260

\bibitem[Isella et al.(2016)]{Isella2016} Isella, A., Guidi, G., Testi, L., et al.\ 2016, Physical Review Letters, 117, 251101

\bibitem[Isella et al.(2018)]{Isella2018} Isella, A., Huang, J., Andrews, S.~M., et al.\ 2018, \apjl, 869, L49

\bibitem[Jin et al.(2016)]{Jin2016} Jin, S., Li, S., Isella, A., Li, H., \& Ji, J.\ 2016, \apj, 818, 76


\bibitem[Johansen et al.(2014)]{Johansen2014} Johansen, A., Blum, J., Tanaka, H., et al.\ 2014, Protostars and Planets VI, 547


\bibitem[Johansen et al.(2009)]{Johansen2009} Johansen, A., Youdin, A., \& Klahr, H.\ 2009, \apj, 697, 1269

\bibitem[Johansen \& Youdin(2007)]{Johansen2007} Johansen, A., \& Youdin, A.\ 2007, \apj, 662, 627
\bibitem[Klahr \& Bodenheimer(2003)]{Klahr2003} Klahr, H.~H., \& Bodenheimer, P.\ 2003, \apj, 582, 869

\bibitem[Li et al.(2001)]{Li2001} Li, H., Colgate, S.~A., Wendroff, B., \& Liska, R.\ 2001, \apj, 551, 874

\bibitem[Li et al.(2000)]{Li2000} Li, H., Finn, J.~M., Lovelace, R.~V.~E., \& Colgate, S.~A.\ 2000, \apj, 533, 1023

\bibitem[Li et al.(2009)]{Li2009} Li, H., Lubow, S.~H., Li, S., \& Lin, D.~N.~C.\ 2009, \apjl, 690, L52

\bibitem[Li et al.(2005)]{Li2005} Li, H., Li, S., Koller, J., et al.\ 2005, \apj, 624, 1003

\bibitem[Liu(2019)]{Liu2019} Liu, H.~B.\ 2019, arXiv:1904.00333 


\bibitem[Liu et al.(2018)]{Liu2018} Liu, S.-F., Jin, S., Li, S., Isella, A., \& Li, H.\ 2018, \apj, 857, 87

\bibitem[Long et al.(2018)]{Long2018} Long, F., Pinilla, P., Herczeg, G.~J., et al.\ 2018, arXiv:1810.06044

\bibitem[Loomis et al.(2017)]{Loomis2017} Loomis, R.~A., {\"O}berg, K.~I., Andrews, S.~M., \& MacGregor, M.~A.\ 2017, \apj, 840, 23

\bibitem[Lovelace et al.(1999)]{Lovelace1999} Lovelace, R.~V.~E., Li, H., Colgate, S.~A., \& Nelson, A.~F.\ 1999, \apj, 513, 805

\bibitem[Mathis et al.(1977)]{Mathis1977} Mathis, J.~S., Rumpl, W., \& Nordsieck, K.~H.\ 1977, \apj, 217, 425

\bibitem[Miranda et al.(2017)]{Miranda2017} Miranda, R., Li, H., Li, S., \& Jin, S.\ 2017, \apj, 835, 118 

\bibitem[Natta et al.(2004)]{Natta2004} Natta, A., Testi, L., Neri, R., Shepherd, D.~S., \& Wilner, D.~J.\ 2004, \aap, 416, 179

\bibitem[Ohtsuki et al.(1990)]{Ohtsuki1990} Ohtsuki, K., Nakagawa, Y., \& Nakazawa, K.\ 1990, \icarus, 83, 205 

\bibitem[Papaloizou et al.(2007)]{Papaloizou2007} Papaloizou, J.~C.~B., Nelson, R.~P., Kley, W., Masset, F.~S., \& Artymowicz, P.\ 2007, Protostars and Planets V, 655

\bibitem[Perez-Becker \& Chiang(2011)]{Perez2011} Perez-Becker, D., \& Chiang, E.\ 2011, \apj, 735, 8

\bibitem[Pinilla et al.(2012)]{Pinilla2012} Pinilla, P., Birnstiel, T., Ricci, L., et al.\ 2012, \aap, 538, A114

\bibitem[Pinilla et al.(2014)]{Pinilla2014} Pinilla, P., Benisty, M., Birnstiel, T., et al.\ 2014, \aap, 564, A51

\bibitem[Pinilla et al.(2016)]{Pinilla2016} Pinilla, P., Flock, M., Ovelar, M.~d.~J., \& Birnstiel, T.\ 2016, \aap, 596, A81

\bibitem[Pinilla \& Youdin(2017)]{Pinilla2017} Pinilla, P., \& Youdin, A.\ 2017, Astrophysics and Space Science Library, 445, 91 

\bibitem[Pinte et al.(2016)]{Pinte2016} Pinte, C., Dent, W.~R.~F., M{\'e}nard, F., et al.\ 2016, \apj, 816, 25

\bibitem[Pinte et al.(2018)]{Pinte2018} Pinte, C., Price, D.~J., M{\'e}nard, F., et al.\ 2018, \apjl, 860, L13

\bibitem[Poppe et al.(2000)]{Poppe2000} Poppe, T., Blum, J., \& Henning, T.\ 2000, \apj, 533, 454

\bibitem[Reg{\'a}ly et al.(2012)]{Regaly2012} Reg{\'a}ly, Z., Juh{\'a}sz, A., S{\'a}ndor, Z., \& Dullemond, C.~P.\ 2012, \mnras, 419, 1701

\bibitem[Ricci et al.(2010b)]{Ricci2010b} Ricci, L., Testi, L., Natta, A., \& Brooks, K.~J.\ 2010b, \aap, 521, A66

\bibitem[Ricci et al.(2012)]{Ricci2012} Ricci, L., Trotta, F., Testi, L., et al.\ 2012, \aap, 540, A6

\bibitem[Ricci et al.(2010a)]{Ricci2010a} Ricci, L., Testi, L., Natta, A., et al.\ 2010a, \aap, 512, A15

\bibitem[Rice et al.(2004)]{Rice2004} Rice, W.~K.~M., Lodato, G., Pringle, J.~E., Armitage, P.~J., \& Bonnell, I.~A.\ 2004, \mnras, 355, 543

\bibitem[Rodmann et al.(2006)]{Rodmann2006} Rodmann, J., Henning, T., Chandler, C.~J., Mundy, L.~G., \& Wilner, D.~J.\ 2006, \aap, 446, 211

\bibitem[Semenov et al.(2003)]{Semenov2003} Semenov, D., Henning, T., Helling, C., Ilgner, M., \& Sedlmayr, E.\ 2003, \aap, 410, 611

\bibitem[Shakura \& Sunyaev(1973)]{Shakura1973} Shakura, N.~I., \& Sunyaev, R.~A.\ 1973, \aap, 24, 337

\bibitem[Sheehan \& Eisner(2018)]{Sheehan2018} Sheehan, P.~D., \& Eisner, J.~A.\ 2018, \apj, 857, 18

\bibitem[Smoluchowski(1916)]{Smoluchowski1916} Smoluchowski, M.~V.\ 1916, Zeitschrift fur Physik, 17, 557

\bibitem[Takeuchi \& Lin(2002)]{Takeuchi2002} Takeuchi, T., \& Lin, D.~N.~C.\ 2002, \apj, 581, 1344

\bibitem[Teague et al.(2018)]{Teague2018} Teague, R., Bae, J., Bergin, E.~A., Birnstiel, T., \& Foreman-Mackey, D.\ 2018, \apjl, 860, L12

\bibitem[Testi et al.(2014)]{Testi2014} Testi, L., Birnstiel, T., Ricci, L., et al.\ 2014, Protostars and Planets VI, 339

\bibitem[Testi et al.(2001)]{Testi2001} Testi, L., Natta, A., Shepherd, D.~S., \& Wilner, D.~J.\ 2001, \apj, 554, 1087

\bibitem[Testi et al.(2003)]{Testi2003} Testi, L., Natta, A., Shepherd, D.~S., \& Wilner, D.~J.\ 2003, \aap, 403, 323

\bibitem[Tsukagoshi et al.(2016)]{Tsukagoshi2016} Tsukagoshi, T., Nomura, H., Muto, T., et al.\ 2016, \apjl, 829, L35

\bibitem[Umebayashi \& Nakano(1981)]{Umebayashi1981} Umebayashi, T., \& Nakano, T.\ 1981, \pasj, 33, 617

\bibitem[van der Marel et al.(2019)]{vanderMarel2019} van der Marel, N., Dong, R., di Francesco, J., Williams, J.~P., \& Tobin, J.\ 2019, \apj, 872, 112 

\bibitem[van der Walt et al.(2011)]{vanderWalt2011} van der Walt, S., Colbert, S.~C., \& Varoquaux, G.\ 2011, Computing in Science and Engineering, 13, 22

\bibitem[Varni{\`e}re \& Tagger(2006)]{Varniere2006} Varni{\`e}re, P., \& Tagger, M.\ 2006, \aap, 446, L13

\bibitem[van Terwisga et al.(2018)]{Terwisga2018} van Terwisga, S.~E., van Dishoeck, E.~F., Ansdell, M., et al.\ 2018, arXiv:1805.03221

\bibitem[Weidenschilling(1977)]{Weidenschilling1977} Weidenschilling, S.~J.\ 1977, \mnras, 180, 57

\bibitem[Wilner et al.(2005)]{Wilner2005} Wilner, D.~J., D'Alessio, P., Calvet, N., Claussen, M.~J., \& Hartmann, L.\ 2005, \apjl, 626, L109

\bibitem[Wilner et al.(2000)]{Wilner2000} Wilner, D.~J., Ho, P.~T.~P., Kastner, J.~H., \& Rodr{\'{\i}}guez, L.~F.\ 2000, \apjl, 534, L101

\bibitem[Youdin \& Goodman(2005)]{Youdin2005} Youdin, A.~N., \& Goodman, J.\ 2005, \apj, 620, 459

\bibitem[Youdin \& Lithwick(2007)]{Youdin2007} Youdin, A.~N., \& Lithwick, Y.\ 2007, \icarus, 192, 588

\bibitem[Zhang et al.(2018)]{Zhang2018} Zhang, S., Zhu, Z., Huang, J., et al.\ 2018, \apjl, 869, L47

\bibitem[Zhu et al.(2019)]{Zhu2019} Zhu, Z., Zhang, S., Jiang, Y.-F., et al.\ 2019, arXiv:1904.02127 


\end{thebibliography}
\end{document}